\documentclass[preprint]{aastex}

\shorttitle{Gigahertz-peaked spectra pulsars}
\shortauthors{Kjak et al.}

\begin{document}

\title{Gigahertz-peaked spectra pulsars and thermal absorption model}
\author{J. Kijak\altaffilmark{1}, R. Basu\altaffilmark{1},  W. Lewandowski\altaffilmark{1}, 
K. Ro\.zko\altaffilmark{1}, M. Dembska\altaffilmark{2}}
\altaffiltext{1}{Janusz Gil Institute of Astronomy, University of Zielona G\'ora, ul. Z. Szafrana 2, PL-65-516 Zielona G\'ora, Poland} 
\altaffiltext{2}{DLR Institute of Space Systems, Robert Hooke Str. 7, 28359 Bremen, Germany}
\email{jkijak@astro.ia.uz.zgora.pl}

\begin{abstract}
We present the results of our radio interferometric observations of pulsars at 
325 MHz and 610 MHz using the Giant Metrewave Radio Telescope (GMRT). We used 
the imaging method to estimate the flux densities of several pulsars at 
these radio frequencies. The analysis of the shapes of the pulsar spectra 
allowed us to identify five new gigahertz-peaked spectra (GPS) pulsars. Using 
the hypothesis that the spectral turnovers are caused by thermal free-free 
absorption in the interstellar medium, we modeled the spectra of all known 
objects of this kind. Using the model, we were able to put some observational 
constrains on the physical parameters of the absorbing matter, which allows us 
to distinguish between the possible sources of absorption.  We also discuss the possible effects 
of the existence of GPS pulsars on future search surveys, showing that the optimal frequency range 
for finding such objects would be from a few GHz (for regular GPS sources) to possibly 
10~GHz for pulsars and radio-magnetars exhibiting very strong absorption.
 \end{abstract}

%\keywords{pulsars: general -- pulsars: individual: J1550$-$5418, J1622$-$4950  -- stars: winds, outflows -- ISM: general}

\section{Introduction}
The radio spectra in pulsars, typically characterized by an inverse power law, 
have been important in understanding the non-thermal origin of radio emission in
pulsars. In a majority of pulsars the spectral nature is described by a steep 
power-law function with an average spectral index of $-$1.8 \citep{maron2000}. In recent years a new class of pulsars have been identified with a distinct spectral nature showing a turnover around 1~GHz \citet{kijak2011b}. These sources, classified the gigahertz-peaked spectra (GPS) pulsars, exhibit the typical power law spectrum at higher frequencies, but their observed
flux decreases with frequency and the corresponding spectral index becomes 
positive at frequencies $<$ 1~GHz. A systematic study was conducted by 
\citet{kijak2007,kijak2011b,kijak2013} and \citet{dembska2014,dembska2015a} where they reported 
nine pulsars and two magnetars to exhibit GPS characteristics. The GPS 
phenomenon appears to be associated with relatively young pulsars found in or 
around peculiar environments such as pulsar wind nebulae (PWNe), H~II regions, etc. This motivated 
\citet{kijak2011b} to suggest that the spectral turnover ($\sim 1$ GHz) is a result of the 
interaction of the radio emission with the pulsars' environments. 

\begin{table}
%\begin{deluxetable*}{cc D{,}{\pm}{3.3}}
%D@{+/-}D}
%\resizebox{\hsize}{!}{
%\begin{minipage}{80mm}
\caption{Observing details.}
\centering
%\startdata
\begin{tabular}{ccc}
%\begin{tabular}{cc D{,}{\pm}{3.3}}
\hline
 & & \\
Obs Date & Phase Calibrator & \multicolumn{1}{c}{Flux}\\
 &  & \multicolumn{1}{c}{({\footnotesize Jy})}\\
\hline
\multicolumn{3}{c}{Day 1, 610 MHz}  \\
 28,29 Dec, 2013 & 1830$-$360 & 15.8$\pm$1.0 \\
 28,29 Dec, 2013 & 1822$-$096 & 6.1$\pm$0.4 \\
 28,29 Dec, 2013 &  1924+334  & 6.4$\pm$0.5 \\
 28,29 Dec, 2013 &  2021+233  & 2.8$\pm$0.2 \\
 28,29 Dec, 2013 & 2047$-$026 & 4.3$\pm$0.3 \\
 & & \\
\multicolumn{3}{c}{Day 2, 610 MHz}  \\
 04,07 Jan, 2014 & 1830$-$360 & 15.7$\pm$0.9 \\
 04,07 Jan, 2014 & 1822$-$096 & 6.2$\pm$0.4 \\
 04,07 Jan, 2014 &  2021+233  & 3.2$\pm$0.3 \\
 04,07 Jan, 2014 & 2047$-$026 & 4.5$\pm$0.3 \\
 & & \\
\multicolumn{3}{c}{Day 1, 325 MHz}  \\
 03,04 Jan, 2015 & 1714$-$252 & 4.3$\pm$0.2 \\
 03,04 Jan, 2015 & 1822$-$096 & 2.7$\pm$0.3 \\
 & & \\
\multicolumn{3}{c}{Day 2, 325 MHz}  \\
 17,18 Jan, 2015 & 1714$-$252 & 4.6$\pm$0.3 \\
 17,18 Jan, 2015 & 1822$-$096 & 3.5$\pm$0.3 \\
% & & \\
\hline
% & & \\
\end{tabular}
\label{tabobs}
%\end{minipage}
%}
%\enddata
\end{table}

The unique binary system of PSR B1259$-$63 and Be star LS 2883 provides a 
window into the GPS phenomenon \citep{kijak2011a}. The spectrum of B1259$-$63
at various orbital phases mimics the spectrum of a GPS pulsar. \citet{kijak2011a} considered two mechanisms that might influence the observed radio 
emission: free-free absorption and cyclotron resonance. 
Most GPS pulsars have no companion and thus there is no direct analogy between them and the binary system. However, the appearance of the GPS in isolated pulsars may be caused, like in the case of PSR B1259-63/LS 2883, by their peculiar environments (like pulsar wind nebulae, supernova remnants filaments). 
The interaction of the radio emission from pulsars with their environments was
investigated in detail by \citet{lewandowski2015a}. They showed that the 
physical properties of certain of environments suggest that thermal absorption 
can cause the observed spectra to turnover at gigahertz frequencies. The 
difference between the binary system and a typical GPS pulsar is that for PSR 
B1259$-$63 the intensity of the effect changes due to variable amount of matter
that the pulsar radiation has to pass through, \citep{dembska2015b}, whereas for the 
isolated pulsar the geometry of the absorber remains static, producing a 
stable, GPS-type spectrum. Similar approach was also used by 
\citet{rajwade2016a} for a selected sample of six pulsars and by 
\citet{basu2016} to explain the apparent variability of the spectra in 
PSR~B1800$-$21.

The statistical studies of the past pulsar search surveys (see for example 
\citealt{bates2013}) indicate that the pulsars exhibiting the GPS phenomenon can
amount up to 10\% of the entire pulsar population. However, the relatively 
small sample of GPS pulsars (11 known cases) is understandable given the pulsar 
spectra are not well studied at low radio frequencies (below 1~GHz). The flux 
measurement becomes particularly challenging in pulsars with suspected GPS 
characteristics. The specialized environments around these pulsars imply that 
they generally have relatively high ($> 200$ pc cm$^{-3}$) Dispersion Measure 
(DM) and their profiles are significantly smeared by interstellar scattering; 
see \citet{lewandowski2013,lewandowski2015b,lewandowski2015c} and \citet{krishna15} for recent studies on scattering. As explained in \citet{dembska2015a}, the traditional flux measurement technique using single dish or 
phased-array telescopes is inadequate to measure the flux in highly scattered 
pulsars leading to gross underestimation of the pulsar flux. Interferometric 
imaging is the only method to securely measure the pulsar flux in such cases. 
Additionally, \citet{basu2016} carried out a detailed comparison between the 
two flux measurement schemes and demonstrated that the interferometric imaging 
is a vastly superior technique to determine the pulsar flux.

The primary objectives of this paper are twofold. Firstly, we have measured 
the flux in a large number of pulsars at low radio frequencies using the 
interferometric methods. This is intended at extending the potential sample of 
GPS pulsars and also verifying the GPS nature in a number of cases as a 
continuation of the studies initiated in \citet{kijak2011b,kijak2013,dembska2014,dembska2015a}. Secondly, we have carried out detailed modeling using the
thermal free-free absorption following the suggestion of \citet{kijak2011a,kijak2011b,kijak2013}. The model fits were carried out for all pulsars exhibiting GPS 
characteristics and expanding the study of \citet{lewandowski2015a} and 
\citet{basu2016}. We used the data fitted models to estimate the physical 
properties of the absorbing electron gas - its density, temperature and 
possible sizes of the absorber.

\section{Observations and data analysis}

We have conducted extensive observations using the Giant Metrewave Radio 
Telescope (GMRT) located near Pune, India. The GMRT consists of an array of 30 
distinct dishes, each with a diameter of 45 meters and  spread 
out over a region of $\sim$ 27 km in a Y-shaped array. The GMRT operates mostly
in the meter wavelengths, between 150 MHz and 1.4 GHz, and is ideally suited to 
measure the pulsar flux at the low radio frequencies and check their spectral
shape. We used the 325 MHz and 610 MHz frequency bands for our studies and 
observed nine pulsars at 325 MHz and eight pulsars at 610 MHz, respectively.
The data were recorded in the interferometric observing mode with each 
frequency band having a bandwidth of 33 MHz spread over 256 spectral channels.
The 610 MHz observations were conducted between December 2013 and January, 
2014 while the 325 MHz observation was carried out in January 2015. Each 
source was observed for roughly one hour during each observing run at both  
frequencies. All observations were repeated twice and the observing sessions were 
separated by a week to account for intrinsic flux variations as well as systematics. 

\begin{table}
%\resizebox{\hsize}{!}{
\begin{minipage}{80mm}
\caption{Flux density measurements   ($S_1$, $S_2$) resulted from the 
observations in two separate observing sessions at 325 MHz and 610 MHz, 
respectively, along with the uncertainties which include calibration errors, 
the rms noise in the maps and fitting errors.$\left<S\right>$ denotes weighted 
mean of all results and its uncertainty for a given pulsar.}
\centering
\begin{tabular}{lccc}
%@{}l  D{,}{\pm}{3.3}  D{,}{\pm}{3.3}  D{,}{\pm}{4.4}@{}}
\hline
 & & &  \\
Pulsar & \multicolumn{1}{c}{$S_1$} & \multicolumn{1}{c}{$S_2$}  & \multicolumn{1}{c}{$\left<S\right>$}\\
& \multicolumn{1}{c}{(mJy)} & \multicolumn{1}{c}{(mJy)} &\multicolumn{1}{c}{(mJy)}\\
%Pulsar & \multicolumn{1}{c}{$S_1$} & \multicolumn{1}{c}{$S_2$}  & \multicolumn{1}{c}{$\left<S\right>$}\\
%& \multicolumn{1}{c}{(mJy)} & \multicolumn{1}{c}{(mJy)} &\multicolumn{1}{c}{(mJy)}\\
\hline

\multicolumn{4}{c}{325 MHz}\\
B1641$-$45 & 140.2$\pm$7.9 & 152.1$\pm$8.5  & 145.7$\pm$8.2\\

J1723$-$3659 & 1.1$\pm$0.3 & 1.4$\pm$0.3 & 1.3$\pm$0.3 \\

B1822$-$14 & \multicolumn{1}{c}{$<$2.5} & 2.4$\pm$0.4  & 2.4$\pm$0.4\\

B1823$-$13 & 1.2$\pm$0.3 & 1.3$\pm$0.2 & 1.3$\pm$0.3\\

B1832$-$06 & 13.6$\pm$0.9 & 16.0$\pm$1.0 & 14.7$\pm$0.9\\

J1835$-$1020 & 1.6$\pm$0.4 & 2.4$\pm$0.3 & 2.1$\pm$0.3 \\

J1841$-$0345 & 1.3$\pm$0.3 & 2.3$\pm$0.6 & 1.5$\pm$0.4\\

J1852$-$0635 & 1.3$\pm$0.3 & 1.8$\pm$0.2 & 1.6$\pm$0.2\\

J1901+0510 & 1.0$\pm$0.3 & 1.2$\pm$0.3 & 1.1$\pm$0.3\\

\multicolumn{4}{c}{610 MHz}  \\

%B1713$-$40 & \multicolumn{1}{c}{$<$0.65} & \multicolumn{1}{c}{$<$0.5}  & \multicolumn{1}{c}{~~~---} \\

J1834$-$0731 & 3.4$\pm$0.5 & 4.5$\pm$0.5 & 4.0$\pm$0.5\\

J1834$-$0812 & \multicolumn{1}{c}{$<$0.85} & \multicolumn{1}{c}{$<$0.95}  &  \multicolumn{1}{c}{~~~---}\\

J1852$-$0635 & 5.2$\pm$0.4 & 5.2$\pm$0.4 & 5.2$\pm$0.4\\

J1856+0245 & \multicolumn{1}{c}{$<$0.75} & \multicolumn{1}{c}{$<$0.85}  &  \multicolumn{1}{c}{~~~---}\\

B1904+06 & 2.8$\pm$0.3 & 3.0$\pm$0.3 & 2.9$\pm$0.3  \\

J1916+0748 & \multicolumn{1}{c}{$<$0.45} & \multicolumn{1}{c}{$<$0.45}  &  \multicolumn{1}{c}{~~~---}\\

J2007+2722 & 1.6$\pm$0.3 & 1.7$\pm$0.2 & 1.7$\pm$0.3 \\

\hline
 & & &  \\
\end{tabular}
\label{tabGMRT1}
\end{minipage}
%}
\end{table}

The observations were carried out using standard interferometric schemes with 
strategically spaced calibrators interspersed with the sources. We recorded 
flux calibrators 3C48 and 3C286 at the start and end of observations for around
8--10 minutes. In addition, a number of phase calibrators, spatially close to 
the pulsars, were selected and observed for 3--4 minutes every hour to account 
for the temporal gain variations in each antenna. The phase calibrators were 
also used to correct for fluctuations in the frequency band. During the 325 MHz
observations two phase calibrators were used, 1714$-$252 and 1822$-$096, while 
the 610 MHz observations utilized five different phase calibrators 1822$-$096, 1830$-$360, 
 1924+334, 2021+233 and 2047$-$026. The imaging analysis was carried
out using the Astronomical Image Processing System (AIPS), similarly to Dembska et
al. (2015a) and \citet{basu2016}. The flux scales of the calibrators 3C286 and 
3C48 were set using the estimates of \citet{perley2013} and used to calculate 
the flux of the different phase calibrators (see Table~\ref{tabobs}). We 
reached noise levels of 0.2--0.5 mJy at 325 MHz ensuring detections of all 
sources with flux in excess of 1.0--2.5 mJy (5$\sigma$ detections). All the 
pulsars observed at 325 MHz could be detected in our observations, the only 
exception was B1822$-$14 where the presence of a nearby strong source increased
the noise levels and the pulsar was below detection limit on the first 
observing session. The noise levels in the maps at 610 MHz were between 
0.1--0.2 mJy ensuring detection limits between 0.5--1 mJy. However, we were 
only able to detect four of the seven sources at 610 MHz with the remaining 
pulsars below the detection limit. The detailed results and implications of 
these measurements are discussed in the subsequent sections.

\subsection{The thermal absorption model}

In this section we present the basic tenets of the thermal free-free 
absorption to model the turnover in the spectra (\citealt{sie73,kijak2011a}). 
To model the spectra we used the approach proposed by \citet{lewandowski2015a} and used by \citet{basu2016}. In this model, the intrinsic pulsar spectrum is assumed to be a single power-law, and to estimate the optical depth we used an approximate formula for thermal free-free absorption \citep{RyLa79,Wil2009}. The observed spectrum can be thus written as:
\begin{equation}\label{eqfit}
S(\nu) = A ~ \left( \frac{\nu}{\nu_{0}}\right)^{\alpha} ~ e^{-B~\nu^{-2.1}} ,
%10 \mathrm{GHz}
\label{turnover}
\end{equation}

\noindent
here $A$ is the intrinsic pulsar flux scaling factor (i.e. the flux density at at the frequency of $\nu_0=10$~GHz), 
$\alpha$ is the intrinsic spectral index of the pulsar, and the frequency $\nu$ is in GHz. The optical depth $\tau$ was expressed by a product of the frequency 
dependent term and the frequency independent parameter  $B = 
0.08235 \times T_{\mathrm{e}}^{-1.35}~\mathrm{EM}$,  where $T_{\mathrm{e}}$ is 
the electron temperature and EM is the Emission Measure (in pc cm$^{-6}$). The best fits to the 
measured spectra were obtained using the Levenberg-Marquardt least squares 
algorithm. Compared to the earlier attempt at modeling of the GPS 
pulsar spectra by \citet{rajwade2016a}, the model we use here is more adequate, since we use a full
3-parameter fit, while they performed a 2-parameter fit, using high frequency 
flux measurements as an ``anchor point'', instead of fitting for the intrinsic pulsar flux amplitude $A$.

Table~\ref{fitparams} shows the best fit values for  parameters $A$, $B$ and
$\alpha$, along with the normalized $\chi^2$ and the resulting peak frequency, 
i.e. the observing frequency at which the model reaches maximum flux. We also included the values of reduced $\chi^2$ for a single power law fit for comparison, and as one can see for all pulsars all but one the thermal absorption model provides significantly better fit to the spectra. The 
uncertainties of the fitted parameters were obtained using 3-dimensional 
$\chi^2$ mapping. The dashed lines in the spectra plots shown in the next section represent the envelopes of all the models that lie within 1$\sigma$ boundary of a best fit model for the 
given spectrum\footnote{for a given spectrum all the models that agree with the
best fit to  $1\sigma$ level will in their entirety lie within the envelope.}

\section{Results}

\begin{figure*}
\begin{tabular}{lr}
{\mbox{\includegraphics[width=80mm,angle=0.]{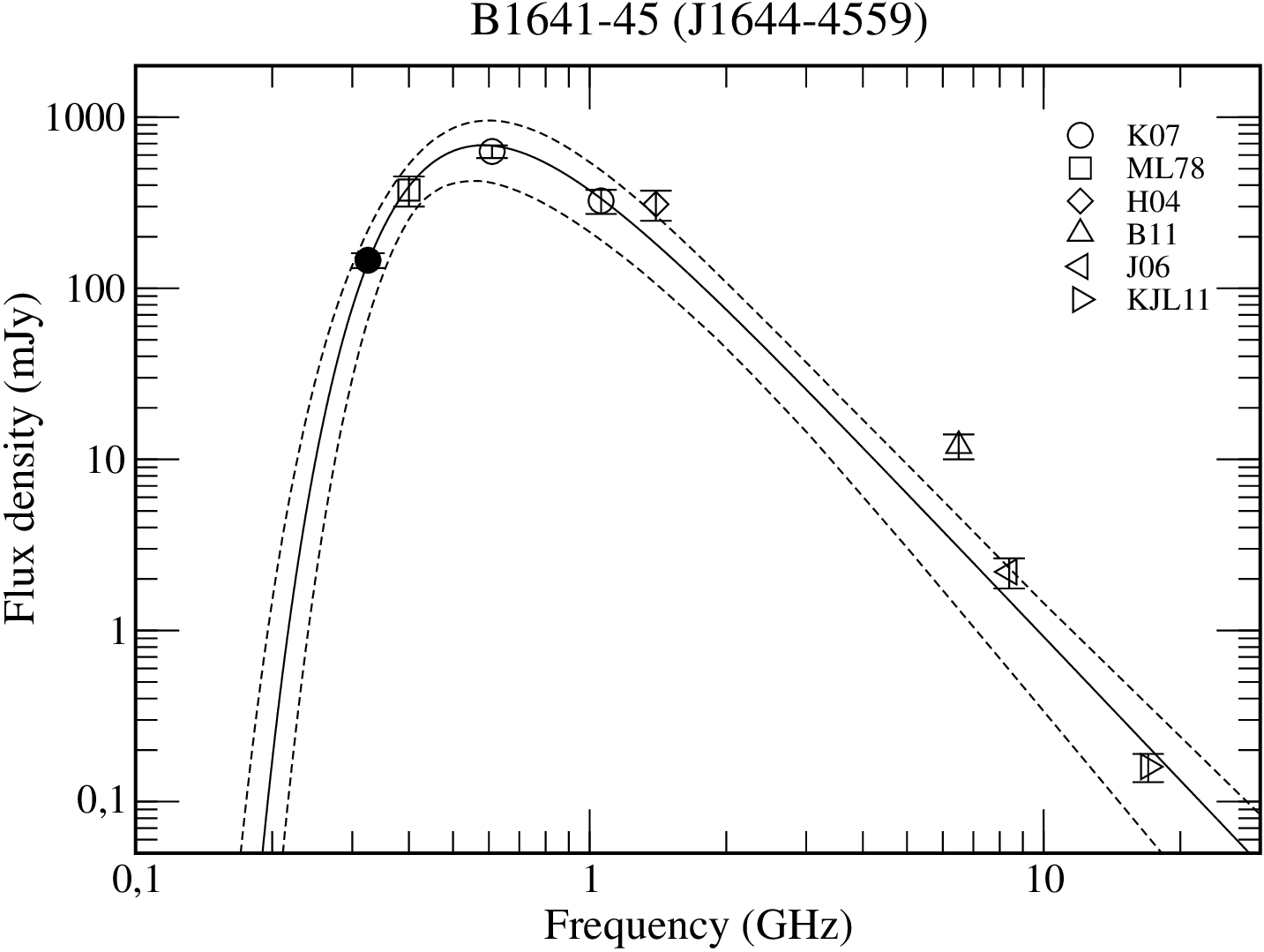}}}&
%{\mbox{\includegraphics[width=80mm,angle=0.]{1713-40.eps}}}\\
{\mbox{\includegraphics[width=80mm,angle=0.]{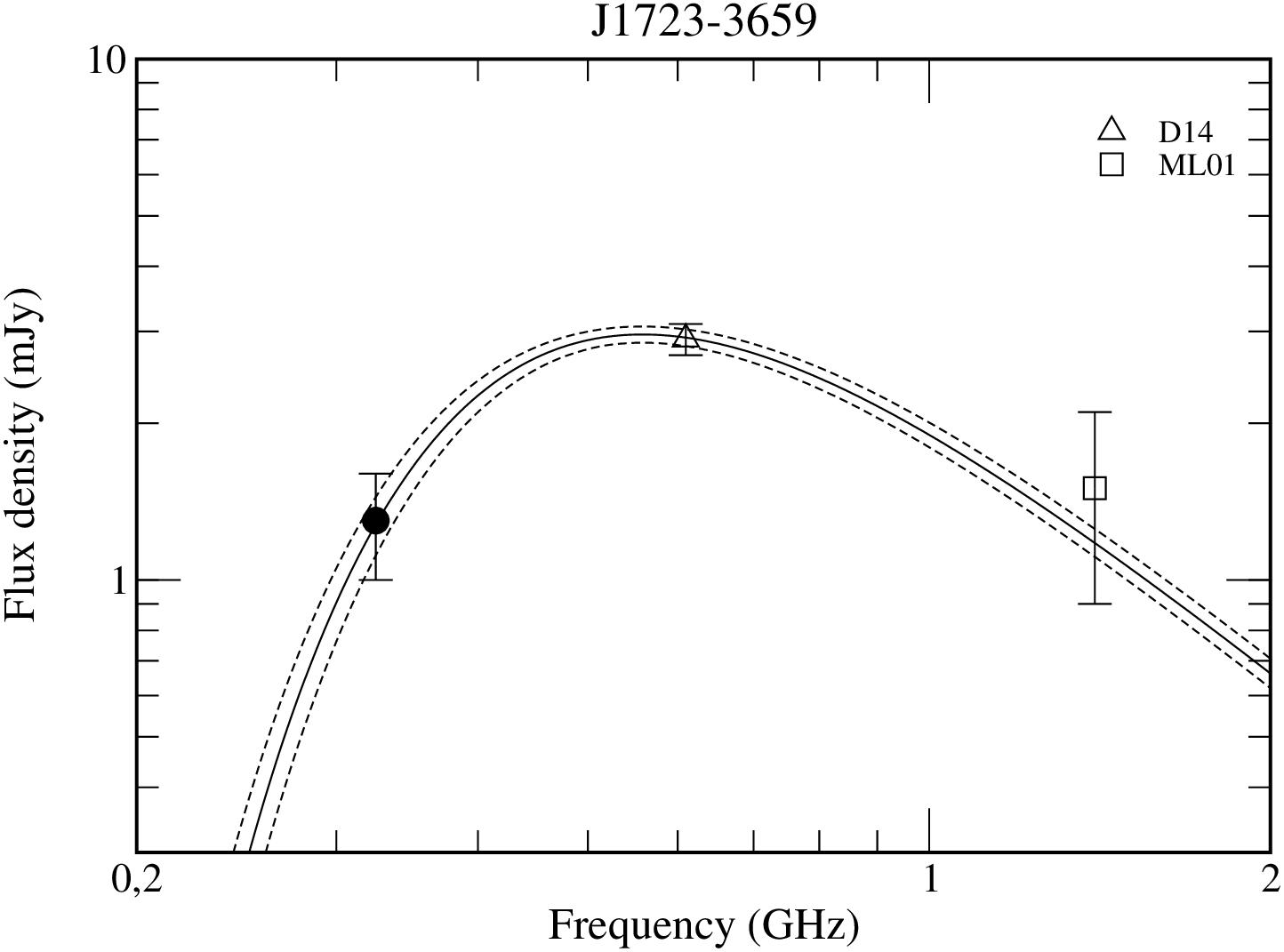}}}\\
{\mbox{\includegraphics[width=80mm,angle=0.]{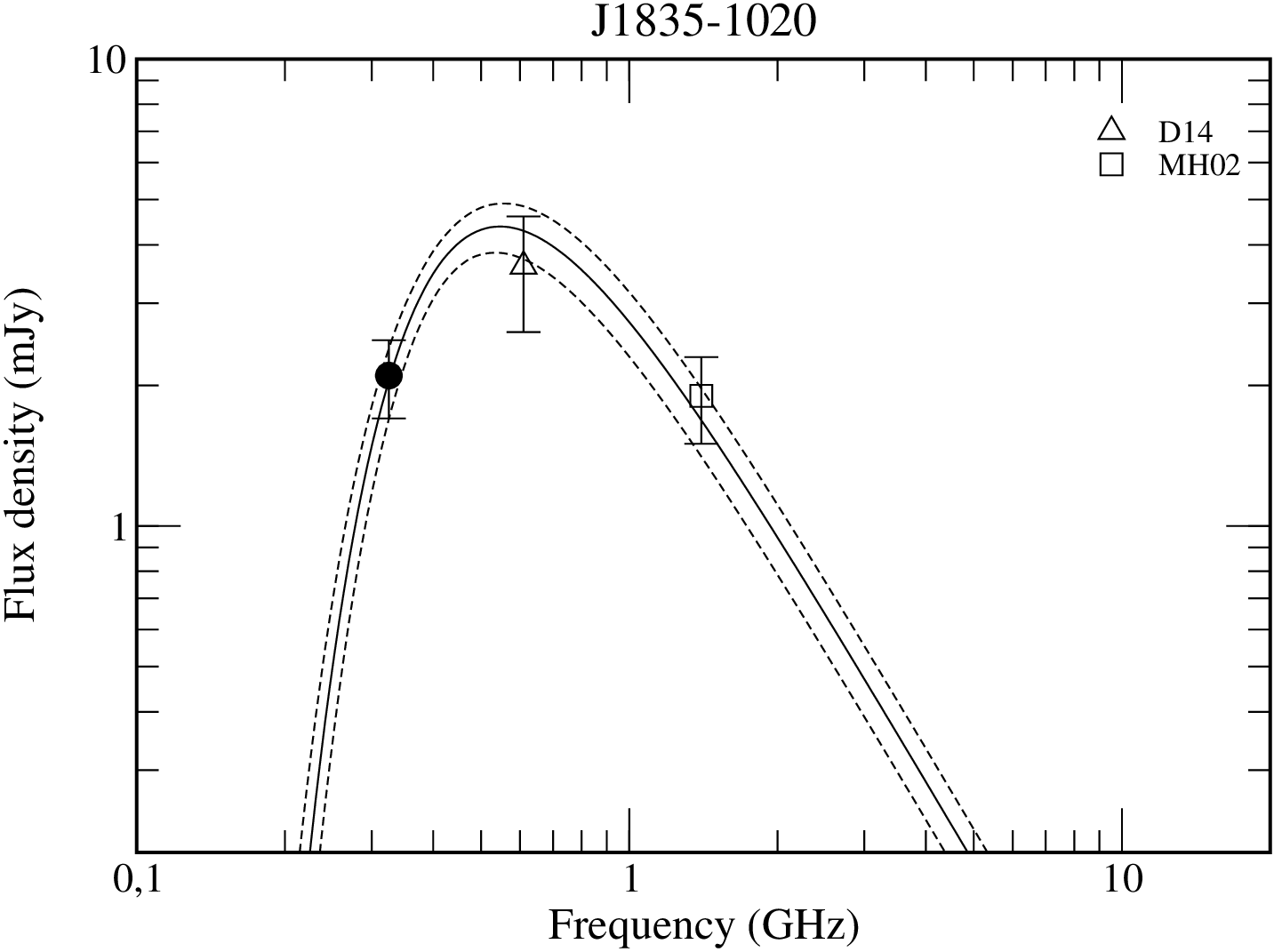}}}&
{\mbox{\includegraphics[width=80mm,angle=0.]{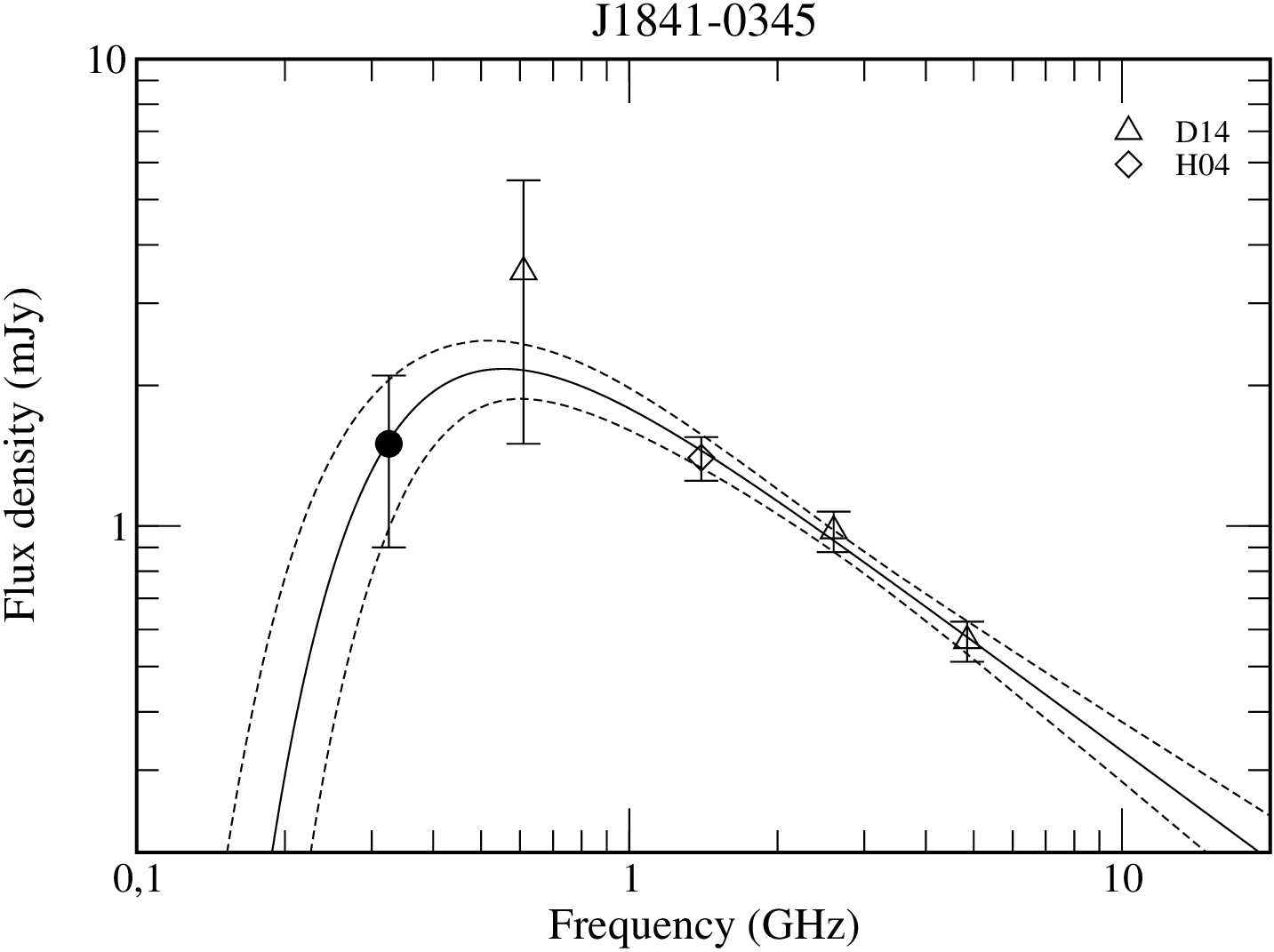}}}\\
{\mbox{\includegraphics[width=80mm,angle=0.]{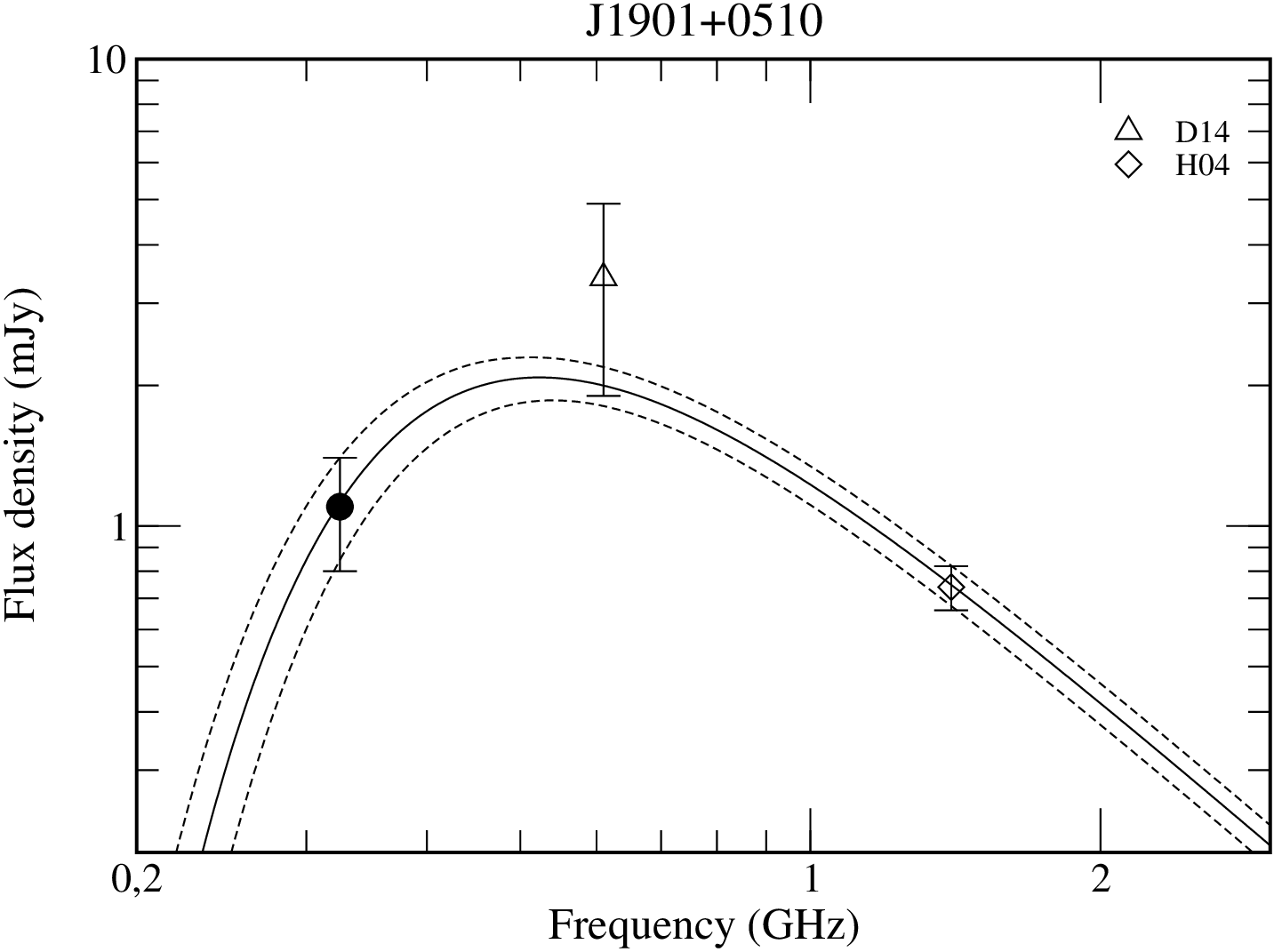}}}&
%\multicolumn{2}{c} {\mbox{\includegraphics[width=87mm,angle=0.]{fig4_b.eps}}}
\end{tabular}
\caption{ New gigahertz-peaked spectra pulsars. The empty circles show flux density from the literature (see References).
Our measurements were denoted by filled circles.  Other values were taken from literature: K07 - \citet{kijak2007}, ML78 - \citet{manchester1978},
H04 - \cite{hobbs2004}, B11 - \citet{bates2011}, J06 - \citet{johnston2006}, KLJ11 - \citet{keith2011}, D14 - \citet{dembska2014}, ML01 - \citet{manchester2001}, MH02 - \citet{morris2002}.
The solid line represents the thermal absorption model fit for the observed data with 1$\sigma$ envelope (dashed lines). 
The fitted parameters are presented in Table~\ref{tabfit}.} 
\label{newGPS}
\end{figure*}

\begin{figure*}
\begin{tabular}{lr}
{\mbox{\includegraphics[width=80mm,angle=0.]{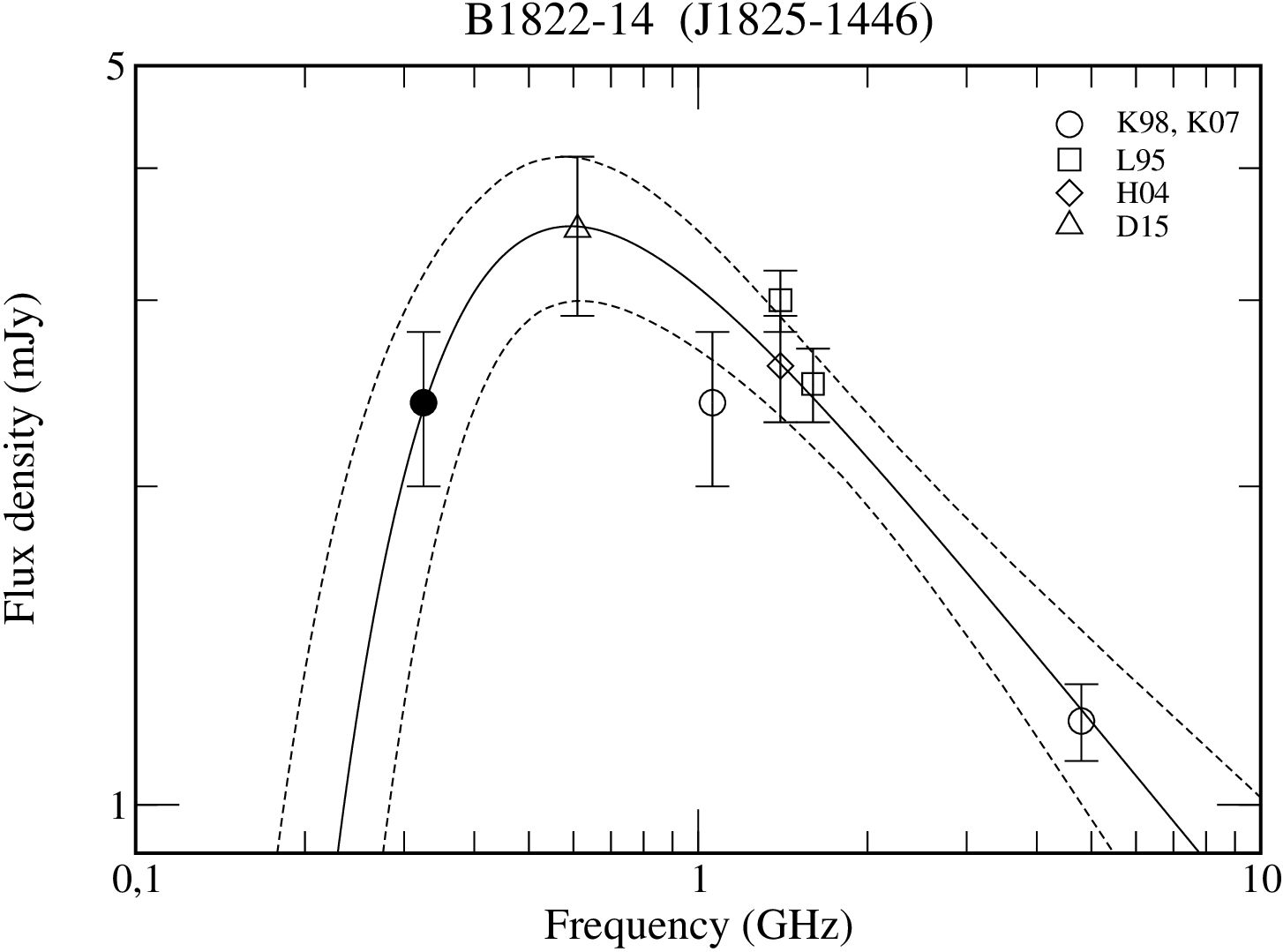}}}&
{\mbox{\includegraphics[width=80mm,angle=0.]{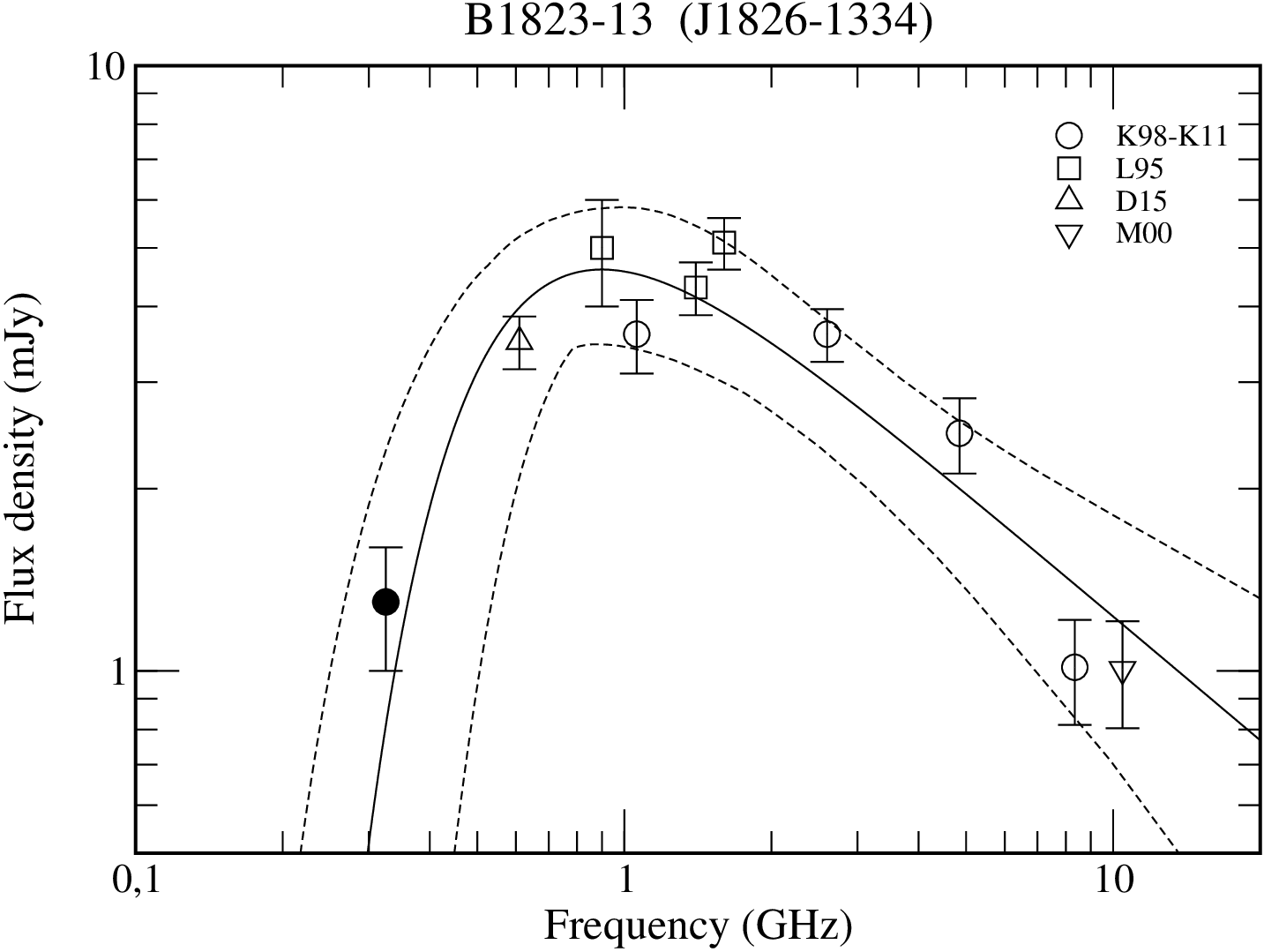}}}\\
{\mbox{\includegraphics[width=80mm,angle=0.]{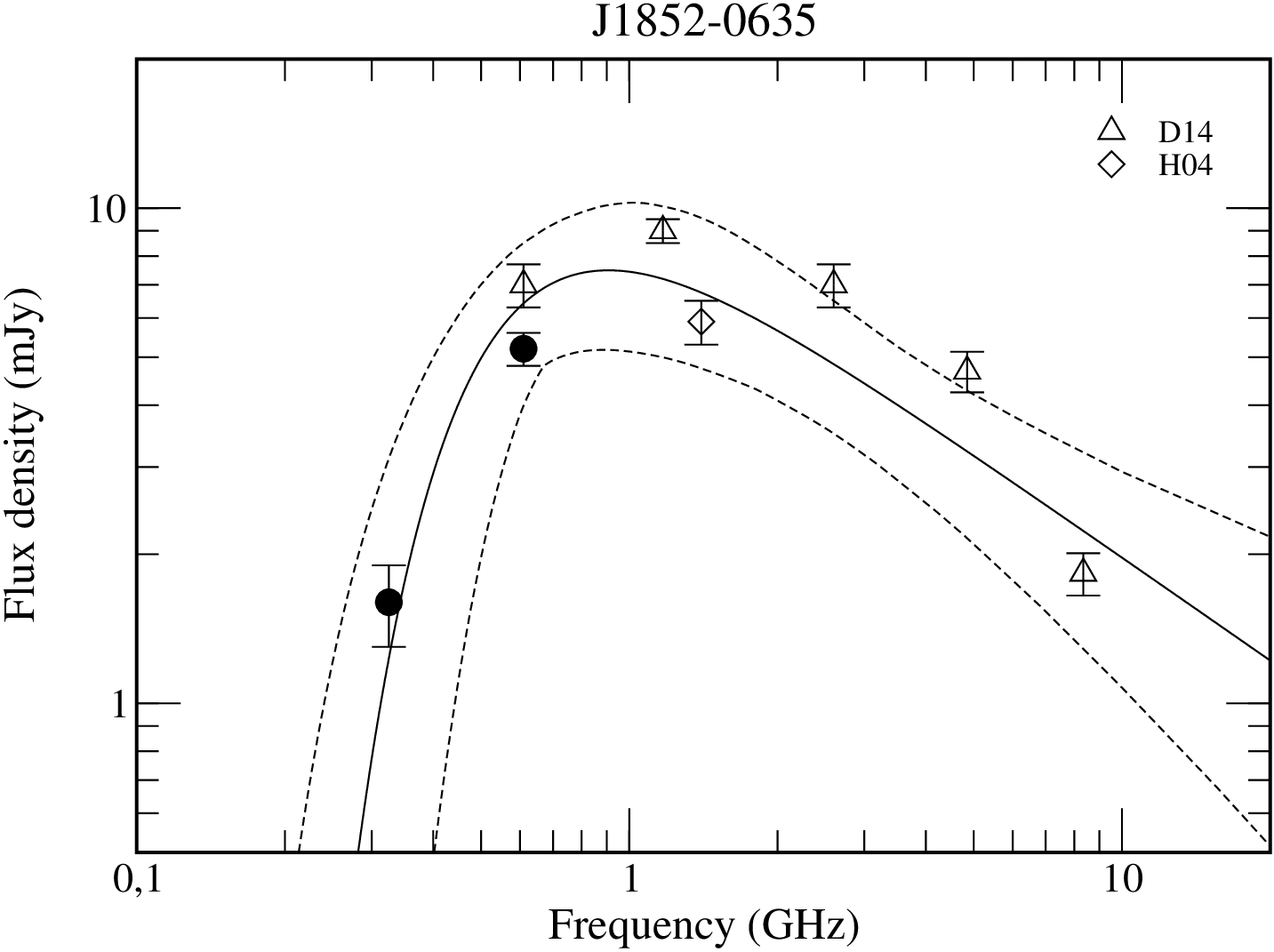}}}&
{\mbox{\includegraphics[width=80mm,angle=0.]{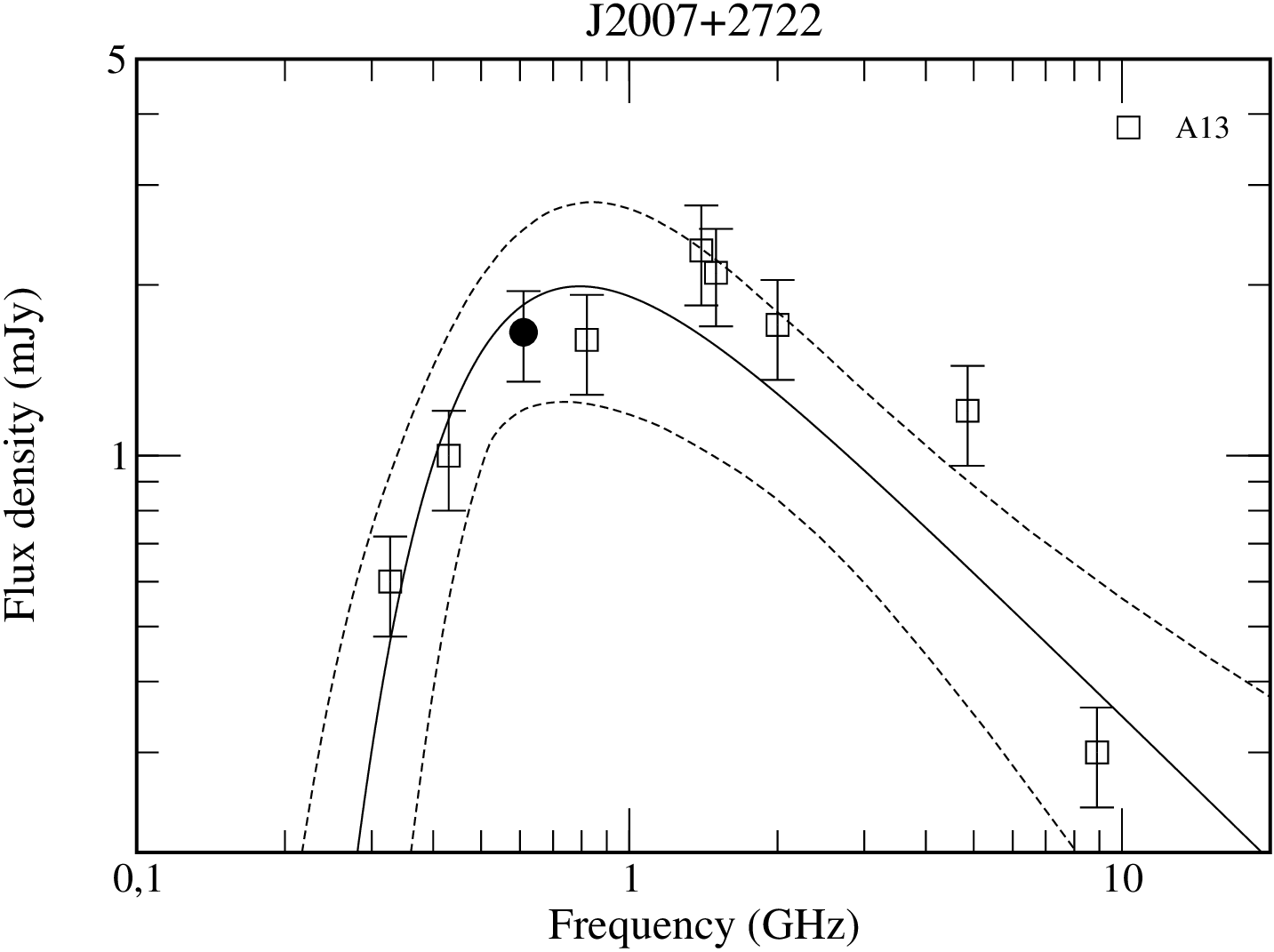}}}
%\multicolumn{2}{c} {\mbox{\includegraphics[width=87mm,angle=0.]{fig4_b.eps}}}
\end{tabular}
\caption{Four confirmed GPS pulsars.  The spectral fits and
errors during fitting shown as solid and broken lines, respectively.
 The fitted parameters and the physical parameters of the absorbing
medium are presented in Tables~\ref{tabfit} and \ref{tabparam}, respectively (see also Fig. \ref{newGPS}).  Filled circles represent our new measurements.  The remaining flux values were taken from literature: K98-K11 - \citet{kijak1998,kijak2004, kijak2007,kijak2011b}, L95 - \citet{lorimer1995}, D15 - \citet{dembska2015a}, M00 - \citet{maron2000}, A13 - \citet{allen2013}, see also Fig.~\ref{newGPS}.  For B1822$-$14 and B1823$-$13 we omitted flux measurements at 610~MHz from \citet{lorimer1995} since they were most likely heavily affected by scattering effects - see discussion in the article text.} 
\label{fig2GPS}
\end{figure*}

\begin{figure}
%\begin{flushleft}
\includegraphics[width=8.2cm]{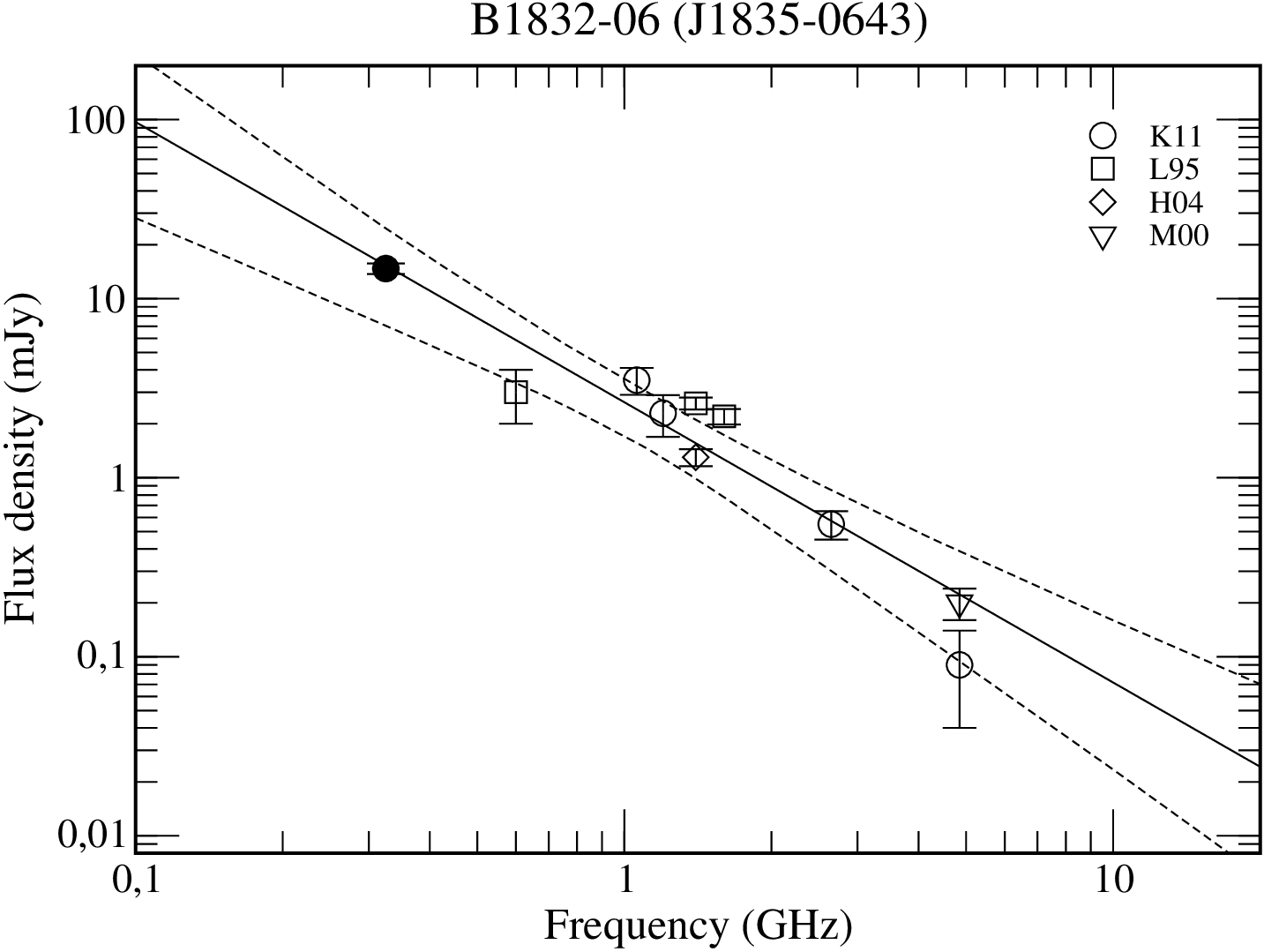}

\includegraphics[width=8.2cm]{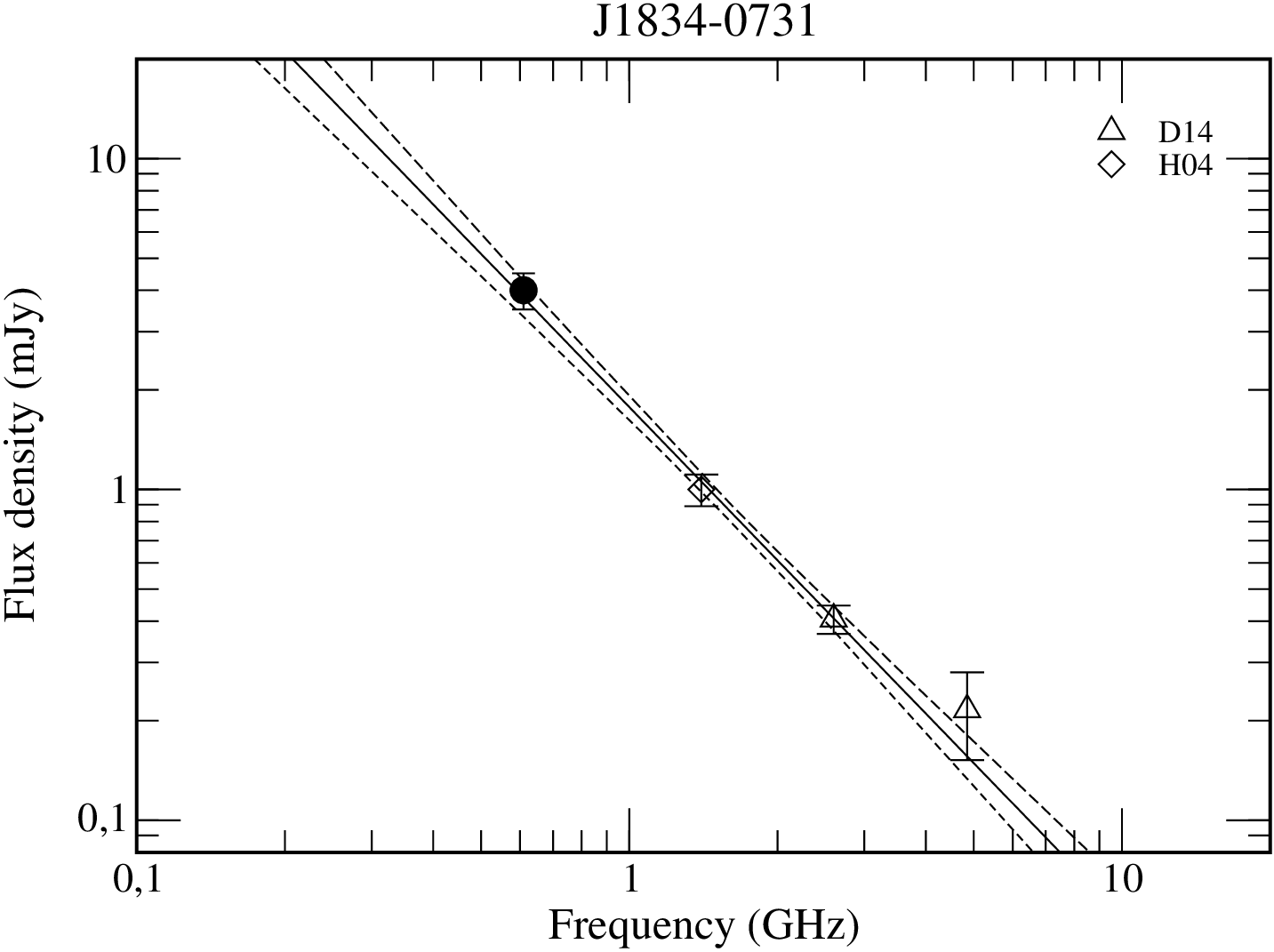}

\includegraphics[width=8.2cm]{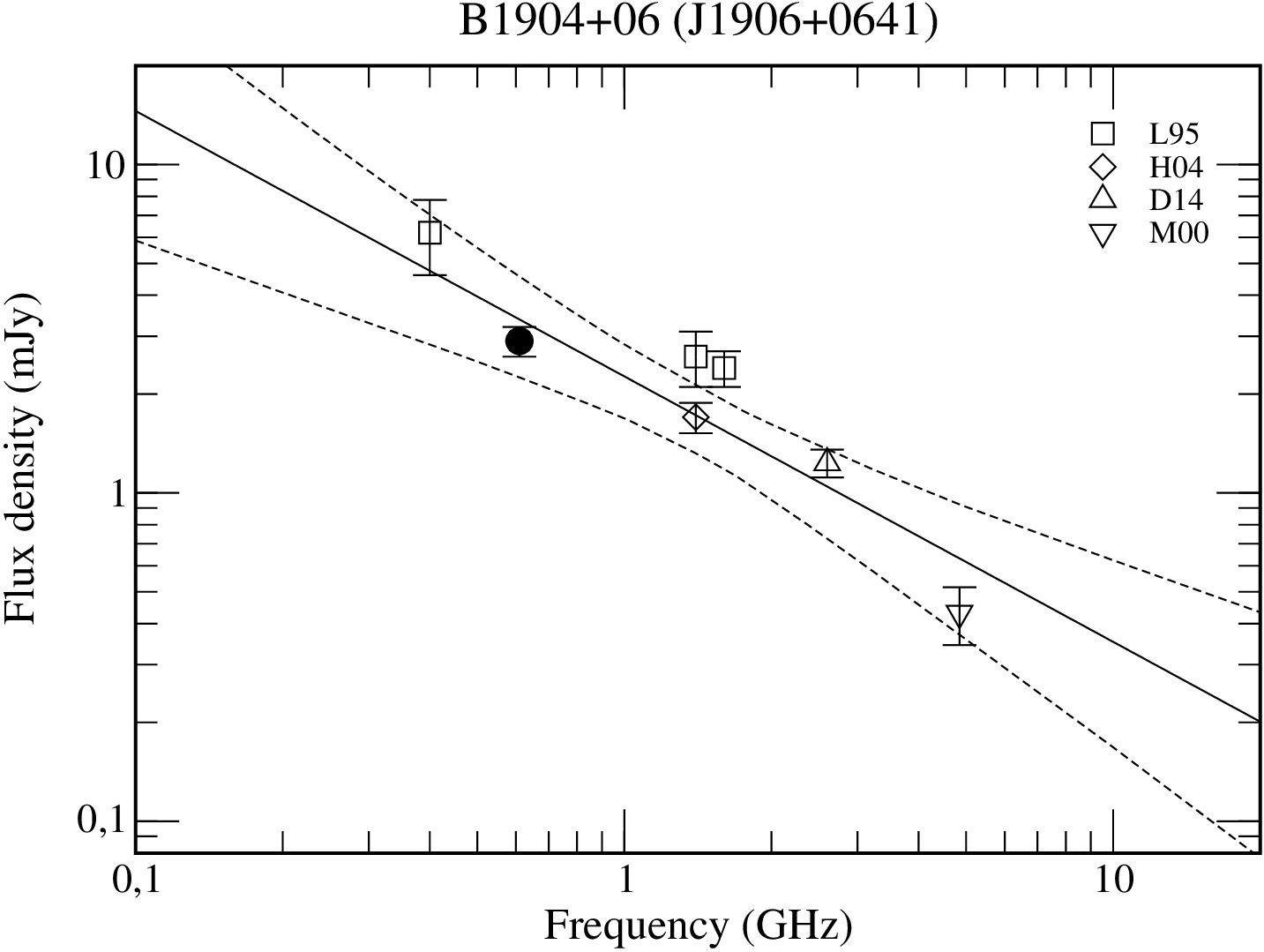}

\caption{The typical pulsar spectra have been modeled using  a simple power law model. Filled circles represent our new measurements, and the remaining flux values were taken from literature, acronyms are explained in previous figures.
The solid line represents the model fit for the observed data with 1$\sigma$ envelope (dashed lines). 
These pulsars B1832$-$06, J1834$-$0731 and B1904+06 have the dispersion measure:
472,  295 and 472  pc~cm$^{-3}$, respectively and are a middle-aged objects: 120, 140 and 1980 kyr, respectively 
(see also Fig. \ref{newGPS}).}
\label{PL}
%\end{flushleft}
\end{figure}

\begin{figure*}
%\begin{flushleft}
\includegraphics[width=8.2cm]{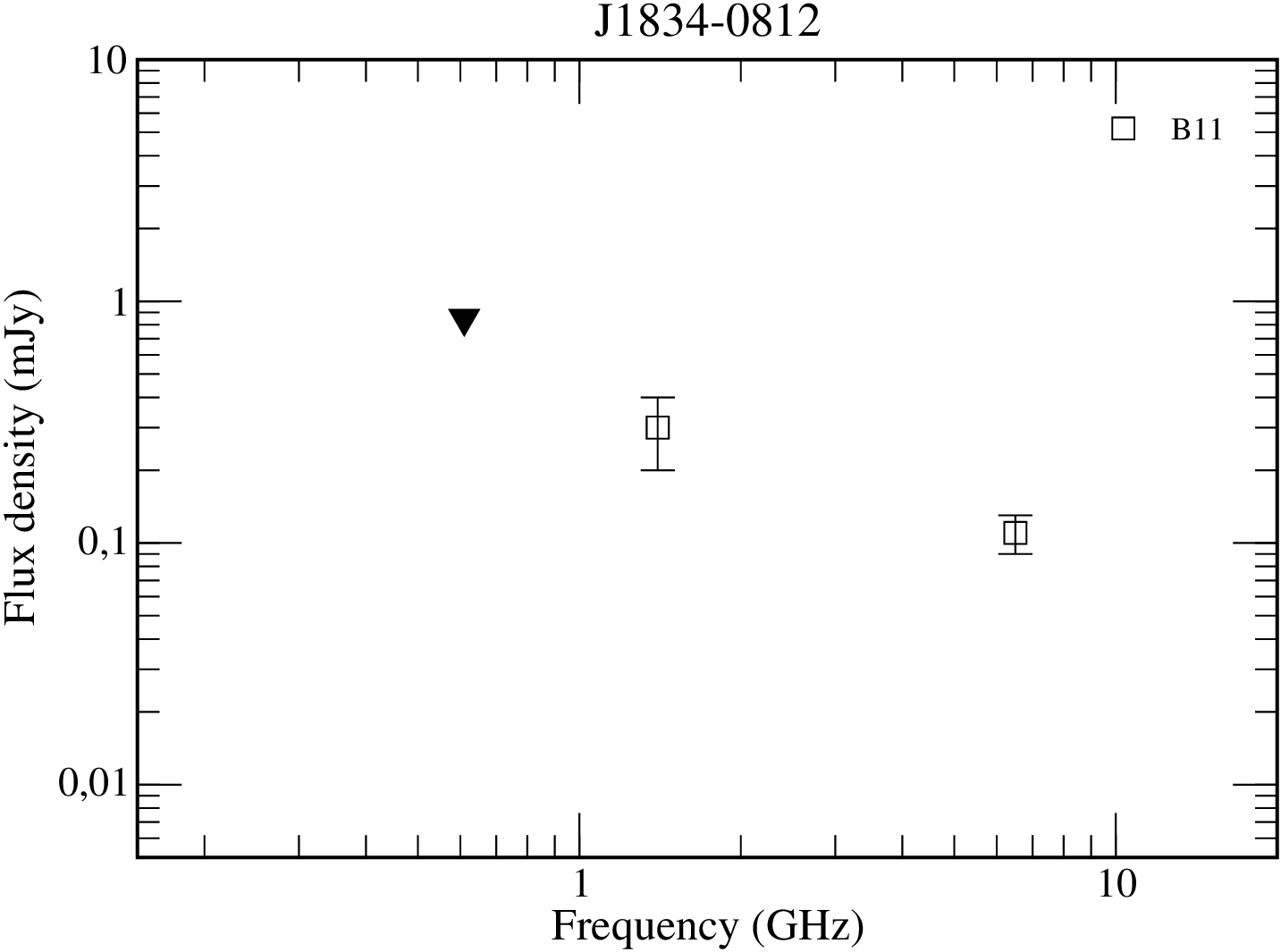}
\includegraphics[width=8.2cm]{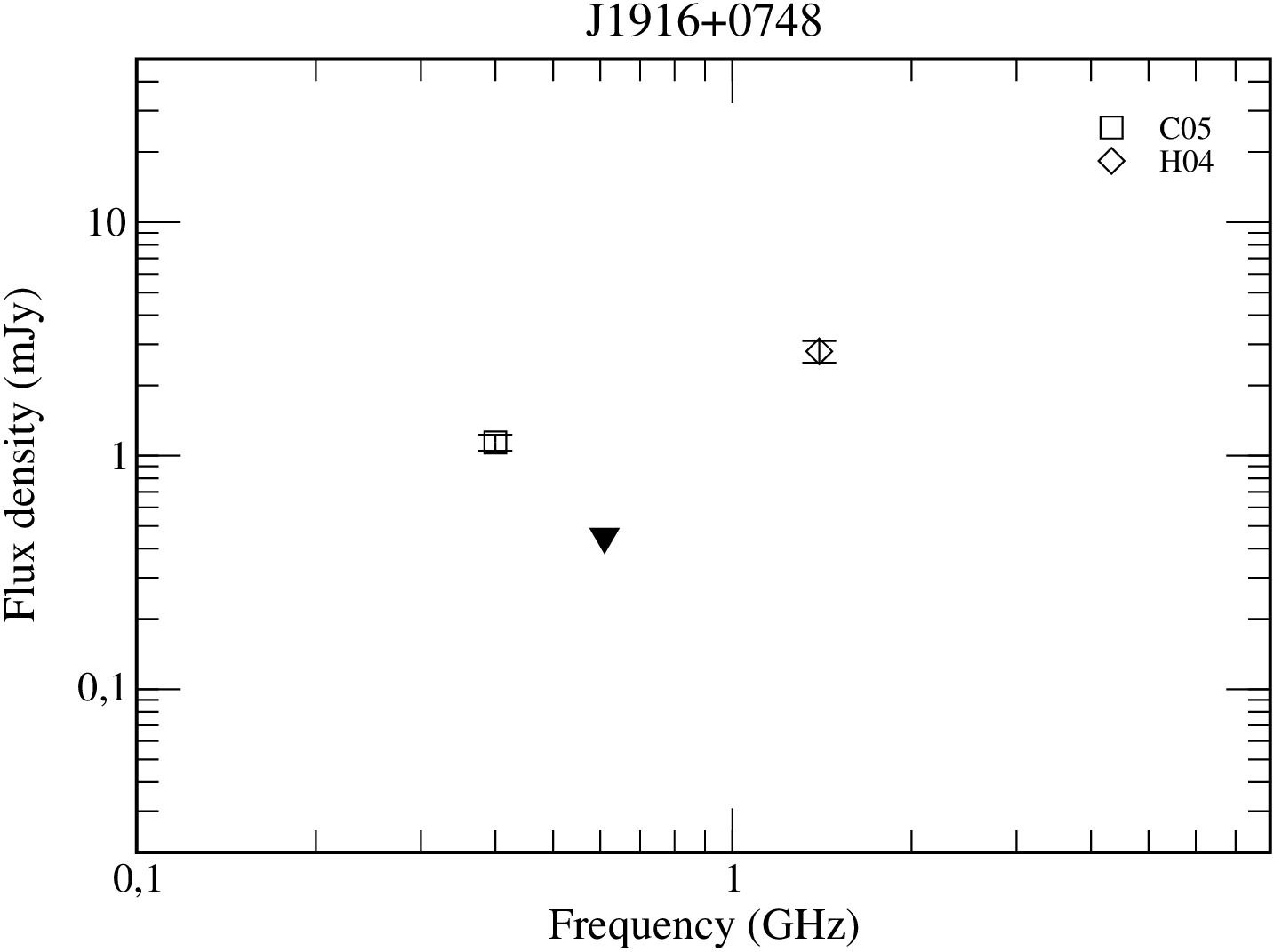}
\caption{Candidates for  GPS pulsars. The inverted triangles show the upper limits from our observations, C05 was taken from \citet{champion2005}, and the remaining flux values were taken from literature, acronyms are explained in previous figures.
PSRs J1834$-$0812 and J1916+0748 have  a characteristic age of 781 and 802 kyr, respectively and 
the dispersion measure DM=1020 and 304~pc~cm$^{-3}$, respectively.}
\label{candidate}
%\end{flushleft}
\end{figure*}

Table~\ref{tabGMRT1} shows the results of our flux density measurements, both for individual observations at both frequencies, as well as the average value.  
As mentioned in the previous section, the pulsars J1834$-$0812, 
J1856+0245 and J1916+0748 were below our detection limit at 610 MHz.
Using the newly acquired data and the previously published flux measurements (see figure captions)
we constructed the spectra for all pulsars that either exhibit the GPS behavior or had been suspected 
to do so before our recent observations. 
The spectra are shown in Figures~\ref{newGPS} to \ref{candidate}, and the new measurements are denoted by filled symbols.
The spectra were divided into 4 groups: the new GPS pulsars (Fig.~\ref{newGPS}), objects for which the GPS feature was confirmed earlier (Fig.~\ref{fig2GPS}), the spectra that resemble a simple power-law (Fig.~\ref{PL}; these pulsars were usually suspected, or previously claimed to show a high frequency turnover), and finally the spectra of two pulsars that still require further investigation (Fig.~\ref{candidate}), which we consider to be promising GPS candidates. In all of these figures except the last one we are also showing the results of spectral modeling fits.

In our analysis we omitted some of the archival measurements for pulsars B1822$-$14 and B1823$-$13 that were published by \citet{lorimer1995}. These measurements were indicated by the authors of that work as suspicious and possibly affected by interstellar scattering. As we mentioned in the Introduction, and also discussed previously in  \citet{kijak2011b} and \citet{dembska2015a} that may lead to severe underestimation of the received flux density, when measured by a standard pulse profile based method \footnote{it is always best to check the profile on which the flux measuremets was based, for example in the EPN Database: http://www.jb.man.ac.uk/research/pulsar/Resources/epn/.}.
The theoretical aspects of the influence of interstellar scattering on the observed pulsar flux were also recently discussed by \citet{geyer2016}. This calls for caution when using archival  (or catalog) flux density measurements, especially at low frequencies, where the effects of scattering-induced flux underestimation will be strongest.

\begin{table*}
\renewcommand*{\arraystretch}{1.5}
\caption{The parameters obtained by modeling of the spectrum of pulsars (see details  in paragraph 3.1). 
The peak frequency $\nu_{\rm p}$ is  the frequency   at which this spectrum displays a maximum flux.\label{fitparams}}
\begin{tabular}{c c c c c c c c} \hline
%&&&&\\
PSR & A & B & $\alpha$ &   $\nu_{\mathrm{p}}{\rm ~(GHz)}$ & $\chi^2$ &  $\chi^2_{\rm PL}$  & \\ 
 %&  &  &  &  & (GHz) \\ 
 \hline
B1054$-$62 & $2.9^{+6.2}_{-2.8}$ & $0.21^{+1.01}_{-0.21}$ & $-1.33^{+0.78}_{-2.34}$ & 0.59 & 12.21 &  14.1 & 5 \\
J1550$-$5418 & $4.16^{+1.02}_{-0.99}$ & $2.6^{+2.7}_{-1.9}$ & $-0.46^{+0.32}_{-0.34}$ & 3.23 & 2.54 &   6.38 &  6 \\
J1622$-$4950 & $10.1\pm {2.6}$ & $4.1^{+3.2}_{-2.3}$ & $-0.54\pm {0.37}$ & 3.73 & 5.09 & 12.9 & 6 \\
B1641$-$45 & $0.92^{+0.53}_{-0.58}$ & $0.43\pm {0.11}$ & $-2.79^{+0.28}_{-0.40}$ & 0.58 & 6.02 & 32.8 & 1 \\ 
J1723$-$3659 & $0.039^{+0.003}_{-0.002}$ & $0.252^{+0.018}_{-0.016}$ & $-1.8$ & 0.56 & 0.30 & 17.6 & 1 \\
J1809$-$1917 & $0.47^{+1.28}_{-0.37}$  &  $2.52^{+4.94}_{-2.52}$  & $-1.34\pm {1.8}$  & 1.93   &  12.1  & 9.9  &  5  \\
B1822$-$14 & $0.77^{+0.25}_{-0.21}$ & $0.104^{+0.055}_{-0.047}$ & $-0.65^{+0.17}_{-0.19}$ & 0.60 & 1.03 & 5.1 & 2 \\
B1823$-$13 & $1.23^{+0.58}_{-0.53}$ & $0.26^{+0.42}_{-0.14}$ & $-0.68^{+0.26}_{-0.42}$ & 0.90 &  3.21 & 17.7 & 2 \\ 
B1828$-$11$^\star$ & $0.010^{+0.008}_{-0.006}$ & $0.85\pm{0.24}$ &  $-2.62^{+0.37}_{-0.48}$  & 0.83   & 1.74 &  17.3  &   7 \\
B1832$-$06  & $0.072^{+0.076}_{-0.044}$ &  --  & $-1.57\pm {0.26}$ & -- & -- &  7.03  & 3 \\
J1834$-$0731 & $0.051^{+0.002}_{-0.003}$  & -- & $-1.54^{+0.13}_{-0.09}$ & -- & -- & 0.66  &  3 \\
J1835$-$1020 & $0.055^{+0.010}_{-0.009}$ & $0.241^{+0.026}_{-0.025}$ & $-1.8$ & 0.55 & 0.78 & 2.6 & 1 \\
J1841$-$0345 & $0.331^{+0.051}_{-0.047}$ & $0.108^{+0.049}_{-0.040}$ & $-0.78\pm {0.11}$ & 0.55 &  0.41 & 2.4 & 1 \\
J1852$-$0635 & $1.97^{+0.97}_{-0.89}$ & $0.27^{+0.37}_{-0.14}$ & $-0.69^{+0.30}_{-0.44}$ & 0.90 & 7.54 & 36.0 & 2 \\
J1901+0510 & $0.024^{+0.0026}_{-0.0025}$ & $0.22^{+0.03}_{-0.02}$ & $-1.8$ & 0.52 & 0.89 & 15.1 & 1 \\
B1904+06  & $0.36^{+0.25}_{-0.17}$ & -- & $-0.79\pm {0.23}$ & -- & -- &  5.35 & 3 \\  
J1907+0918 & $0.0011^{+0.0132}_{-0.0011}$ & $0.43^{+0.65}_{-0.34}$ & $-2.8^{+1.4}_{-2.1}$ & 0.59 &  4.36 & 13.7  & 5 \\
J2007+2722 & $0.35^{+0.21}_{-0.19}$ & $0.25^{+0.22}_{-0.13}$ & $-0.85^{+0.34}_{-0.41}$ & 0.79 & 2.74 & 10.2 & 2 \\
\hline
 \multicolumn{8}{l}{Note: the number in the last unmarked column indicates the number of Figure in which the spectrum,}\\
 \multicolumn{8}{l}{along with the fitted model, is plotted. $\star$ -- is not considered as GPS (see section 3.4). }
\end{tabular}
\label{tabfit}
\end{table*}

\subsection{New and confirmed GPS pulsars}

Figure~\ref{newGPS} shows the spectra of five pulsars for which the GPS effect
is presented for the first time. These objects are characterized by the 
relatively high DM values ($>190$~pc~cm$^{-3}$), except PSR~J1841$-$0345 which
has DM = 56~pc~cm$^{-3}$. Most of the pulsars are relatively young $\tau <$ 
10$^{6}$ year (see Table~\ref{tabparam}, the basic pulsar parameters were taken from the ATNF Catalog\footnote{Available at {\tt http://www.atnf.csiro.au/people/pulsar/psrcat/},  \citet{manchester2005}}).

For pulsars J1723$-$3659, J1835$-$1020 and J1901+0510 our new measurements were only the third in their respective spectra. Having three data points makes it impossible to model the spectrum with three parameters. In the case of these three pulsars, we have limited the number of fitted parameters by an additional assumption that the pulsar's intrinsic spectrum has a spectral index of $-1.8$ (the average value for the non-recycled pulsar population, see \citealt{maron2000}), and estimated 
only  $A$ and $B$ parameters from the fits.

Figure~\ref{fig2GPS} presents the spectra of four objects which were classified as GPS pulsars by~\citet{kijak2011a,kijak2011b,dembska2014} and \citet{allen2013}. The objects were included in our sample to confirm the shape of the spectra and put better constrains on the thermal absorption model by adding new measurements at low observing frequencies (where the free-free absorption manifests itself the strongest). In all four cases the new measurements confirm the earlier claims of high frequency turnovers.

\subsection{Objects with typical spectra and GPS candidate pulsars}

The pulsars whose spectra are shown in Figure \ref{PL} were suspected to show the GPS phenomenon in our earlier studies and hence were included in our observing sample. However, our new measurements (together with the previously published flux density values) clearly show that pulsars B1832$-$06, J1834$-$0731 and B1904+06 exhibit a regular power-law spectra down to 300~MHz.

Figure \ref{candidate} presents spectra of two pulsars we attempted to measure in our observations but only obtained upper limits for their flux density at 610~MHz. We still believe that these objects are good candidates for GPS pulsars.  PSR~J1834$-$0812 exhibits a very high dispersion measure (DM=1020~pc~cm$^{-3}$). For PSR J1916+0748, the upper limit for the flux density clearly suggests a positive spectral index, however we can not explain why our limit is lower than the earlier measured flux at 400~MHz.

\subsection{Radiomagnetars and other GPS pulsars}

\begin{figure}

\includegraphics[width=82mm,angle=0.]{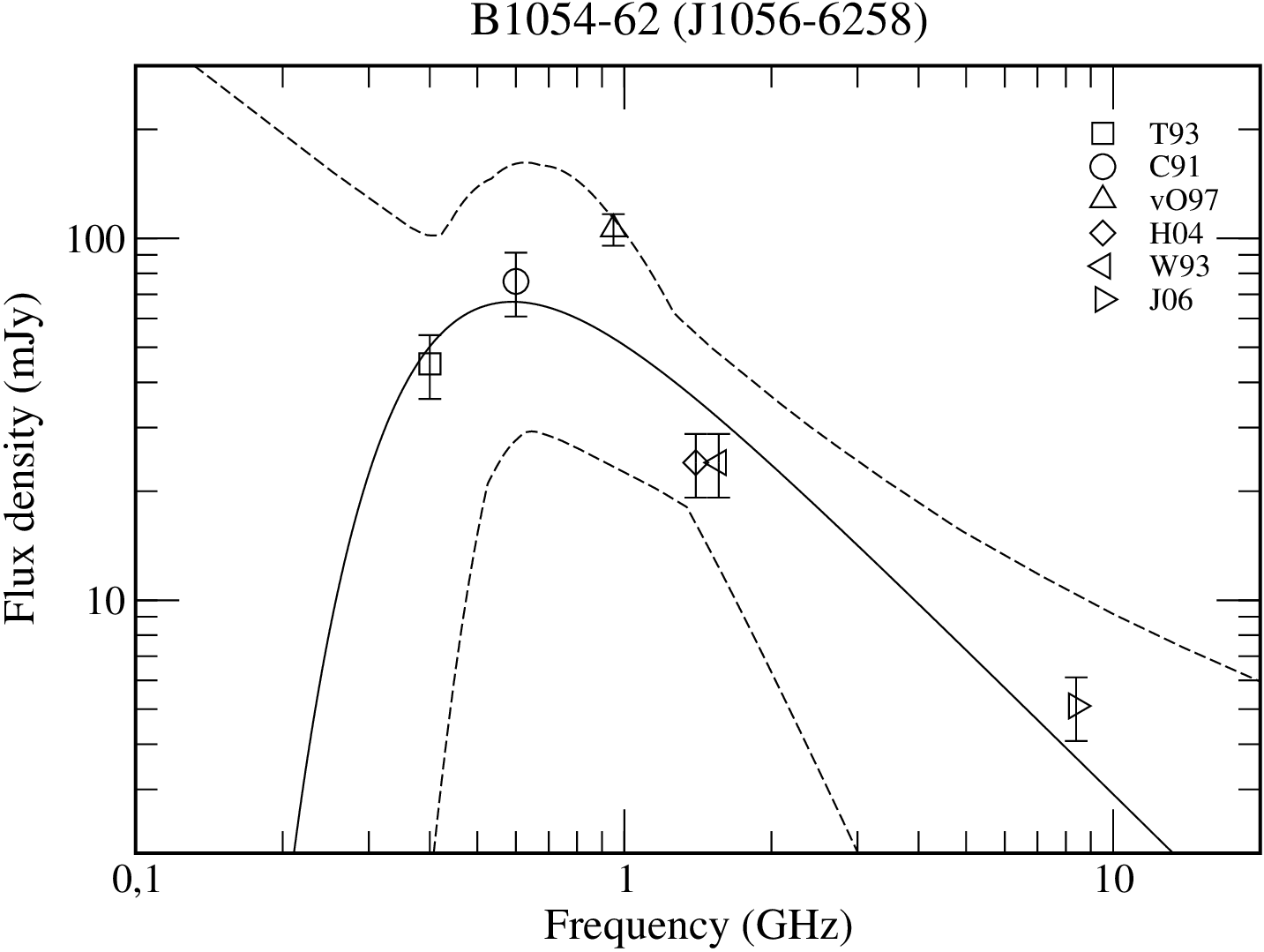}

\includegraphics[width=82mm,angle=0.]{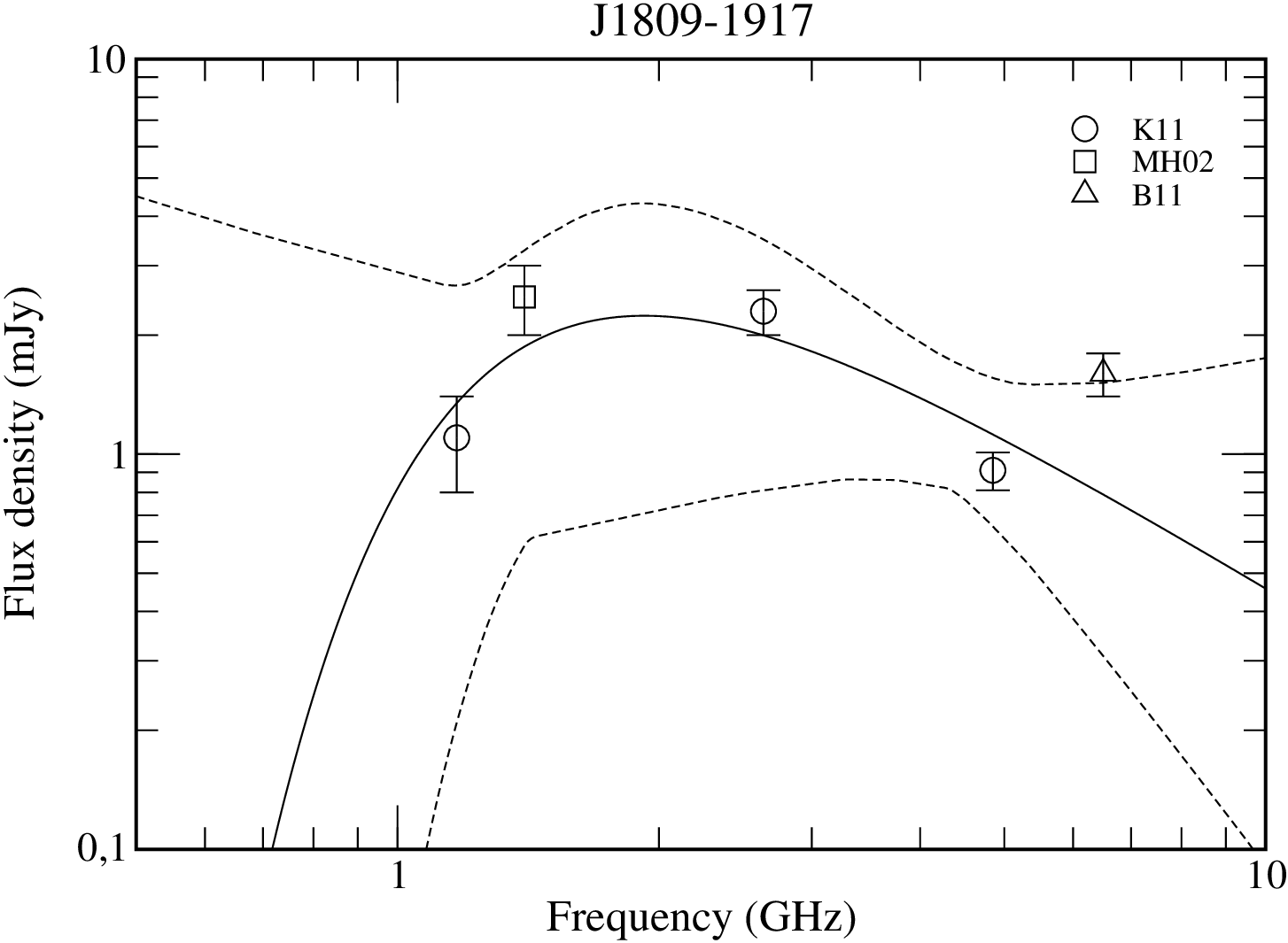}

\includegraphics[width=82mm,angle=0.]{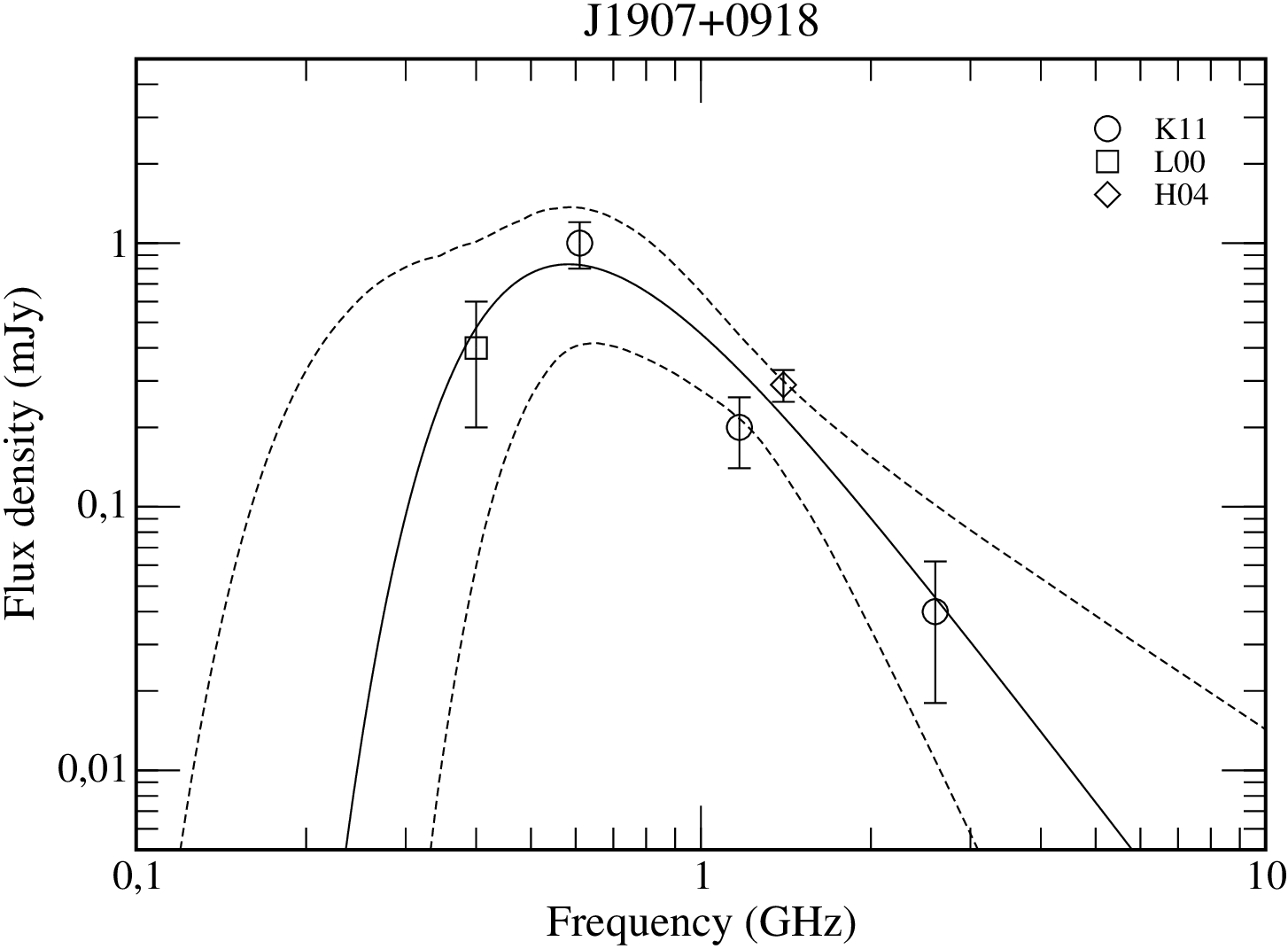}

\caption{Other GPS pulsars with the absorption model fits. Data taken from: T93 - \citet{taylor1993}, C91 - \citet{costa1991}, vO97 - \citet{ommen1997}, W93 - \citet{wu1993}, L00 - \citet{lorimer2000}, and the remaining acronyms are explained in previous figures. } 
\label{fig5}
\end{figure}

\begin{figure*}
\begin{tabular}{lr}
{\mbox{\includegraphics[width=80mm,angle=0.]{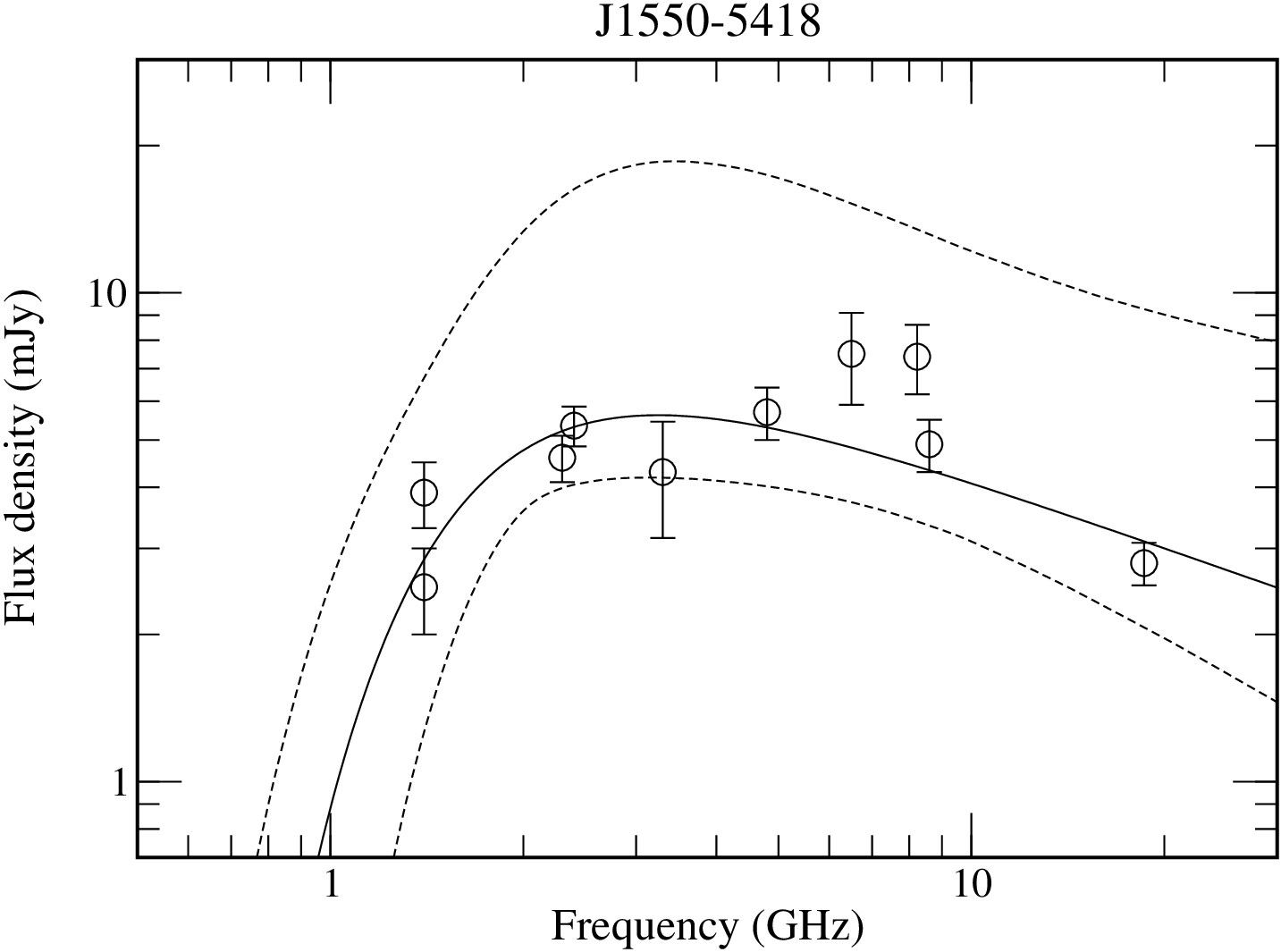}}}&
{\mbox{\includegraphics[width=80mm,angle=0.]{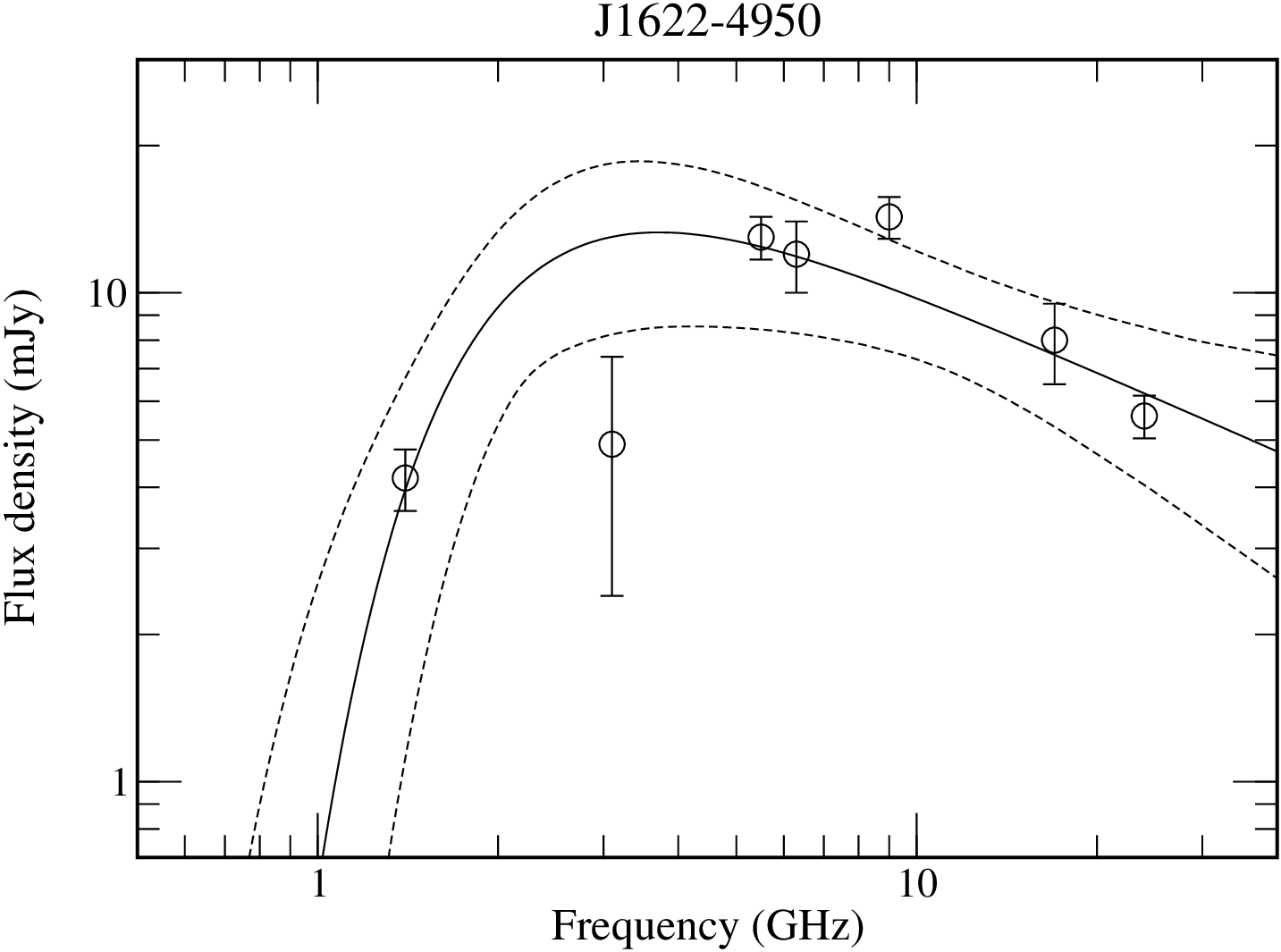}}}
\end{tabular}
   \caption{The absorption model fits for the observed data in two radio-magnetars (see \citealt{kijak2013}). Data are taken from 
   \citet{camilo2007,camilo2008} and \citet{bates2011} for PSR J1550$-$5418 and from \citet{levin2010,levin2012}, \citet{keith2011}, and 
   \citet{anderson2012} for PSR J1622$-$4950.}
  \label{fig6}
\end{figure*}

Figures~\ref{fig5} and \ref{fig6} present the spectra of the GPS pulsars and 
radio magnetars that were published earlier \citet{kijak2007,kijak2011a,kijak2011b,kijak2013}.
We did not add any new data points to these spectra, however, for completeness,
we decided to apply the thermal absorption model to these spectra as well. The 
results of our fits are included in Tables~\ref{fitparams}.
In the case
of PSR~B1054$-$62 there are some results that allow for a power law spectrum to within a $1\sigma$ level. This is indicated  by the shapes of the $1\sigma$ envelopes, and the fact that the uncertainty in  $B$ parameter (see Table~\ref{fitparams}) extends down to include the value of $B=0$ within $1\sigma$ level, which means no absorption. 
Similar is the case of PSR~J1809$-$1917, the only pulsar in Table~\ref{fitparams}, for which the reduced $\chi^2$ from a power-law fit is lower than for the thermal absorption fit. Compared to the spectrum that was shown earlier in \citet{kijak2011b} we added  a measurement from \cite{bates2011} at 6.5~GHz. The large discrepancy between this measurement and our previous 5~GHz measurement causes the fit to be less conclusive. This is because in our modeling the high frequency part of the spectrum is primarily used to ascertain the slope of the intrinsic pulsar spectrum. Hence the discrepancy between two high frequency measurement translates to large uncertainties, as we can not exclude the possibility that the intrinsic spectrum is relatively flat. We decided, however to include this pulsar as a GPS source - the low frequency part of the spectrum clearly indicates a turnover, as the 1170~MHz flux is almost 2.5 times smaller than the 1.4~GHz flux.

\citet{lewandowski2015a} used the thermal absorption model to explain the observed GPS-like evolution of the Sgr~A* radio magnetar spectrum (PSR J1745$-$2900). Here we show the spectra of another two radio magnetars that were pointed out by \citet{kijak2013} to show GPS characteristics. The results of our modeling clearly show that the high frequency turnovers in the spectra are undeniable, since - as it is indicated by the shape of $1\sigma$  envelope - no single power-law model will agree with our results to within $1\sigma$ uncertainty.

\subsection{Break or turnover in the spectrum of the PSR B1828$-$11}

\begin{figure}
%\begin{flushleft}
\includegraphics[width=8.2cm]{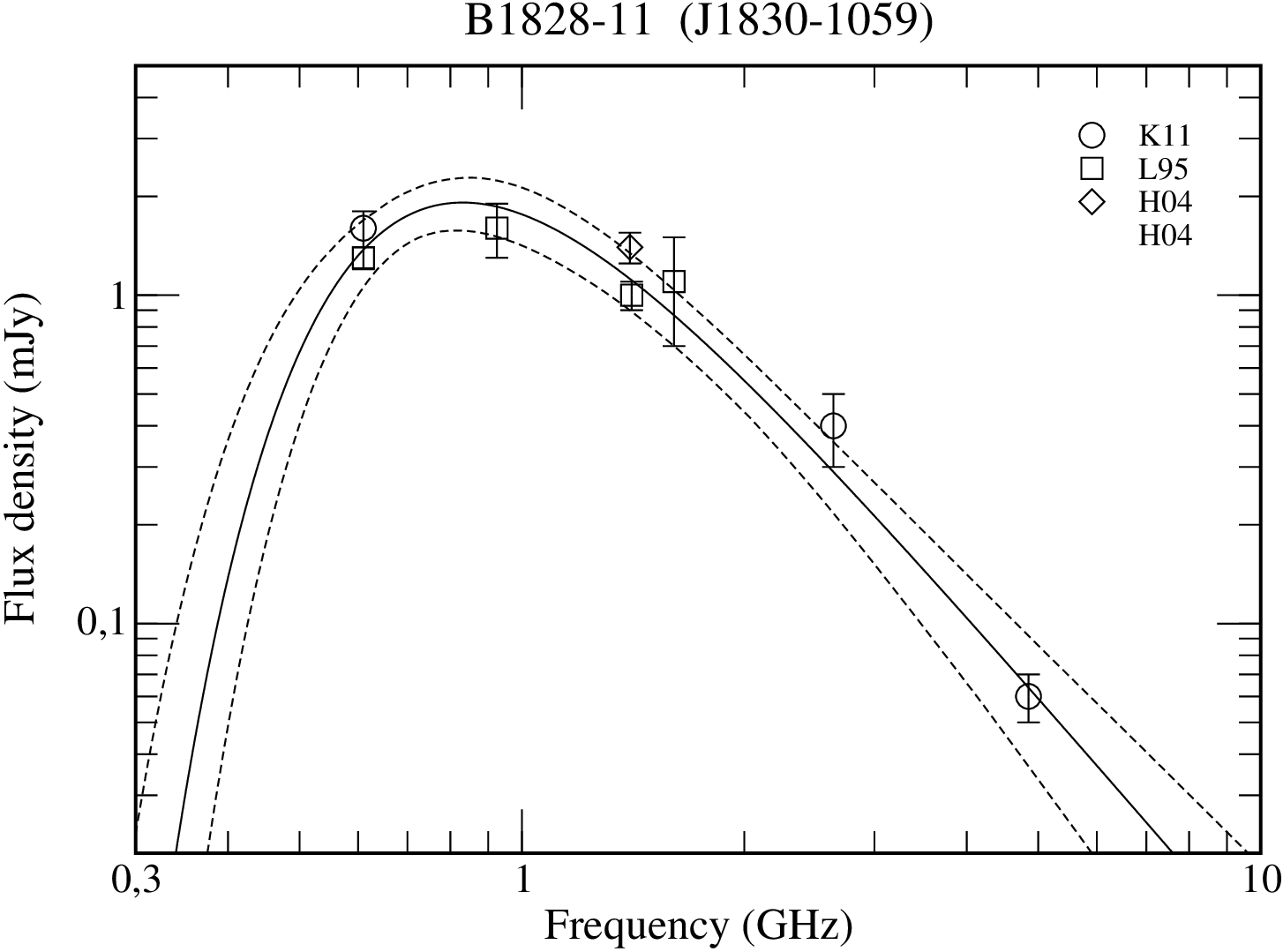}
\caption{The spectrum of PSR B1828$-$11. 
The acronyms are explained in previous figures. The curve represents our fits to the data using the function given 
in Eq.~(\ref{turnover}). }
\label{b1828}
%\end{flushleft}
\end{figure}

In Fig.~\ref{b1828} we present the spectrum of PSR~B1828$-$11. This pulsar was previously studied by \citet{kijak2011b}, and was classified as a flat-normal, or in other words, a broken spectrum, with the spectral index of $-2.4$ in the high frequency range, and 0.5~in the low frequency range, below 1.2 GHz. 
Despite having no new observations for the source, we decided to apply our thermal absorption model to its spectrum (see Table~\ref{tabfit} for the results of our fit). 
We believe that the results of our modeling show that the broken spectra of pulsars can be, at least in some of the cases, explained with the absorption model. The lack of a spectral turnover in such a case may be the result of the limited range of frequencies over which the pulsar flux density measurements were done. To test that hypothesis detailed observations at lower frequencies are necessary.

\section{Discussion}

The interferometric imaging technique is a much superior technique to estimate 
the pulsar flux (see \citealt{basu2016}), and provides the only means to estimate 
the flux of pulsars which are affected by scattering (see discussion in \citealt{dembska2015a}). 
Given the robustness of our flux measurements, we were able
to estimate the low frequency spectra in several pulsars and found five new 
cases of gigahertz-peaked spectra: PSR~B1641$-$45, PSR~J1723$-$3659, 
PSR~J1835$-$1020, PSR~J1841$-$0345 and PSR~J1901+0510. In addition, we also
verified the GPS phenomenon in another four pulsars. The spectral turnover in
fifteen GPS pulsars  was successfully explained 
using the thermal absorption model (see Table \ref{tabparam}). In the remainder of this section we explore
the physical implications of these results.

The spectra of six GPS pulsars were previously modeled using the thermal absorption hypothesis by 
\citet{rajwade2016a}. Their models were using our data from \citet{kijak2007, kijak2011b} and \citet{dembska2014}. For five of these sources we are showing a model based on spectra in a wider frequency range, as we added new flux density measurements, mainly at low frequencies. This, in conjunction  with using a full three parameter model, allows us to obtain much more reliable estimates of the absorbers physical parameters.

\subsection{Thermal absorption as the source of spectral turnovers}

The idea of the thermal free-free absorption as the source of the low frequency spectral turnovers in pulsars was first proposed by \citet{sie73} and was proposed to explain the GPS spectra of pulsars and magnetars by \citet{kijak2011a,kijak2013}.  \citet{lewandowski2015a} showed that indeed the peculiar environments of some pulsars may provide sufficient amounts of absorption to cause the spectra to turnover at gigahertz frequencies. Until now we were aware of the low frequency turnovers (around 100 MHz), and the gigahertz-peaked spectra in which the peak frequency was reported to be around 1~GHz (see \citealt{kijak2011a,kijak2011b,kijak2013,dembska2014,dembska2015a}), however, the results we present here show rather a continuous range of peak frequencies (see Table~\ref{fitparams}), and not merely bi-modal. Also, owing to the use of an actual physical model, instead of purely morphological attempts (using for example \citealt{kuzmin2000}, log-parabolic fits), the peak frequencies we obtained are lower than the values estimated earlier. This is to be expected, since the spectral profile of the absorption will be always asymmetric, with the peak shifted  towards lower frequencies.

The fact that the peak frequencies in the pulsar spectra affected by thermal absorption cover a wide range of values should come as no surprise, since in reality both the ISM and the immediate pulsar surroundings can show a wide, and roughly continuous range of physical parameters relevant to this phenomenon. Depending on the amount of absorption, which in itself is bound to the physical properties of the absorbing matter (electron density, temperature and its extent along the line of sight), one can expect that different configurations will cause the spectra to peak at different frequencies - below 300~MHz for pure ISM, around 1~GHz for pulsars obscured by dense supernova remnant (SNR) filaments, but also the entire range between these two possibilities should be covered by ionized clouds - regions that have electron densities higher than neutral ISM, but not as high as the compact clumps of matter in supernova remnants.

\begin{figure}
\includegraphics[width=82mm,angle=0.]{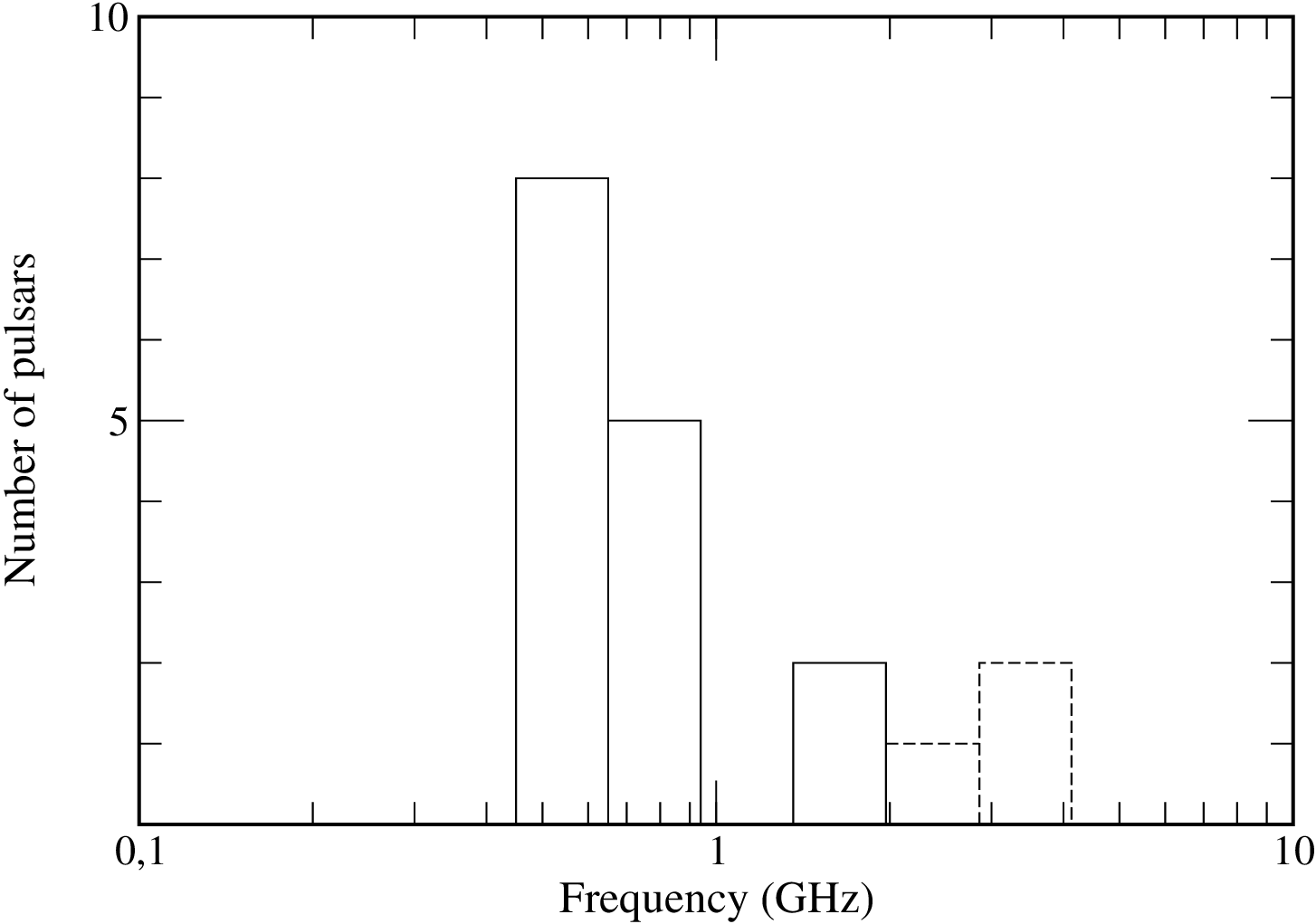}
\caption{Histogram showing the distribution of GPS pulsar/radio magnetar peak frequencies (see also Table~\ref{tabparam}).
The dashed boxes represent the peak frequencies of radio-magnetars.}
\label{fig8}
\end{figure}

\subsection{The absorber physical parameters}

\begin{table*}
{\scriptsize
\caption{The physical parameters of the absorbers derived from the thermal absorption model fits, along with the basic relevant pulsar/magnetar parameters  (taken from the ATNF catalog). The remarks column indicates definite or possible associations of the neutron star with a supernova remnant nebula (SNR), a pulsar wind nebula (PWN), an H II region, or an unidentified X-ray source from the HESS catalog.}
\begin{tabular}{c c c c c c c c c} \hline
PSR & DM & Age & Size & n$_{\mathrm{e}}$ & EM & T & Remarks &  \\ 
 & $\left(\rm{pc~cm}^{-3}\right)$ & (kyr) & (pc) & $\left(\rm{cm}^{-3}\right)$ &  
$\left(10^4~\rm{pc~cm}^{-6}\right)$ &   (K) &  & \\ \hline

B1054$-$62  & 320 & 1870 & 0.1 & $1601 \pm 3$ & $25.648 \pm 0.096$ & $5080^{+18100}_{-3780}$ & H~II (BBW 328) &  \\
 & & & 1.0 & $160.1 \pm 0.3$  & $2.56648 \pm 0.0096$ & $923^{+3282}_{-686}$ & & 5\\
 & & & 10.0 & $16.01 \pm 0.03$ & $0.25648 \pm 0.00096$ & $179^{+596}_{-125}$ & & \\ \hline

J1550$-$5418  & 830 & 1.41& 0.1 & $4150 \pm 250$ & $172.2 \pm 20.8$ & $3210^{+2710}_{-2030} $ & radiomagnetar & \\
  & && 1.0 & $415 \pm 25$ & $17.22 \pm 2.08$ & $583^{+493}_{-369}$ &  SNR & 6 \\
 & & & 10.0 & $41.5 \pm 2.5 $ & $1.7222 \pm 0.208$ & $106^{+90}_{-67}$ & & \\ \hline

J1622$-$4950  & 820 &  4.03& 0.1 & $4100 \pm 150$ & $168.1 \pm 12.3$& $2280^{+1460}_{-1060}$ & radiomagnetar &  \\
 & & & 1.0 & $410\pm 15$ & $16.81 \pm 1.23$ & $414^{+265}_{-193}$  & SNR, PWN?, H~II & 6 \\
 & & & 10.0 & $41.0 \pm 1.5$ & $1.681 \pm 0.123$ & $75^{+48}_{-35}$ & & \\ \hline

B1641$-$45  & 479 &  359& 0.1 & $2394 \pm 4$ & $57.31 \pm 0.19$ & $5413^{+1040}_{-922}$  & HESS &  \\
  & & & 1.0 & $239.4 \pm 0.4$ & $5.731 \pm 0.019$ & $983^{+190}_{-167}$ & (near Westerlund 1 & 1 \\
 & & & 10.0 & $23.94 \pm 0.004$ & $0.57 \pm 0.0019$ & $179^{+34}_{-30}$  &  globular cluster)&\\ \hline

J1723$-$3659  & 254 & 401& 0.1 & $1271.0 \pm 1.5$ & $16.154 \pm 0.038$ & $3150^{+172}_{-153}$  & & \\
 & & & 1.0 & $127.10 \pm 0.15$ & $1.6154 \pm 0.0038$ & $572^{+31}_{-28}$ & & 1 \\
 & & & 10.0 & $12.710 \pm 0.015$ & $0.16154 \pm 0.00038$ & $103.9^{+5.7}_{-5.0}$ & & \\ \hline

 J1809$-$1917  & 197 & 51.3& 0.1 & $986 \pm 2$ & $9.712 \pm 0.039$ & $267^{+36}_{-36}$ & X:bow shock PWN & \\
  & & & 1.0 & $98.55 \pm 0.2$ & $0.9712 \pm 0.0039$ & $48.5^{+6.6}_{-6.6}$ & SNR?, H~II?, HESS & 5 \\
  & & & 10 & $9.855 \pm 0.02$ & $0.09712 \pm 0.00039$ & $8.8^{+1.2}_{-1.2}$ & & \\ \hline

  B1822$-$14 & 357 & 195& 0.1 & $1785 \pm 25$ & $31.86 \pm 0.89$ & $10100^{+4130}_{-3590}$ &  SNR, H~II & \\
 & & & 1.0 & $178.5 \pm 2.5$ & $3.186 \pm 0.089$ & $1830^{+751}_{-652}$ & & 2 \\
 & & & 10.0 & $17.85 \pm 0.25$ & $0.3186 \pm 0.0089$ & $332^{+136}_{-118}$ & & \\ \hline

B1823$-$13  & 231 & 21.4& 0.1 & $1155 \pm 5$ & $13.34 \pm 0.12$ & $2670^{+3250}_{-1070}$ &  X:PWN &  \\
 & & & 1.0 & $115.5 \pm 0.5$ & $1.334 \pm 0.012$ & $485^{+589}_{-194}$ & HESS & 2 \\
 &  & &10.0 & $11.55 \pm 0.05$ & $0.1334 \pm 0.0012$ & $88^{+107}_{-35}$ & & \\ \hline

  B1828$-$11 & 161 & 107& 0.1 & $808 \pm 1$ & $6.52 \pm 0.016$ & $660^{+150}_{-144}$ & &  \\
  & && 1.0 & $80.8 \pm 0.1$ & $0.6521 \pm 0.0016$& $120^{+27}_{-26}$ & & 7 \\
 & & & 10.0 & $8.08  \pm 0.01$ & $0.06521 \pm 0.00016$ & $21.6^{+4.9}_{-4.8}$ & & \\ \hline

  J1835$-$1020  & 114 & 810&  0.1 & $568.5 \pm 4.5$ & $3.232 \pm 0.051$ & $988^{+89}_{-86}$ & & \\
  & && 1.0 & $56.85 0.45$ & $0.3232 \pm 0.0051$ & $180^{+16}_{-16}$ & & 1 \\
 & & & 10.0 & $5.685 \pm 0.045$ & $0.03232 \pm 0.0051$ &  $32.6^{+2.9}_{-2.8}$ & & \\ \hline
 
   J1841$-$0345  & 194 & 55.9& 0.1 & $971.6 \pm 0.3$ & $9.4401 \pm 0.0058$ & $3960^{+1340}_{-1080}$ & H~II? & \\
  & & & 1.0 & $97.16 \pm 0.03$ & $0.94401 \pm 0.00058$ & $720^{+244}_{-196}$ & & 1 \\
 & & & 10.0 & $9.716 \pm 0.003$ & $0.094401 \pm 0.000058$ & $131^{+44}_{-36}$ & & \\ \hline

  J1852$-$0635  & 171 & 567& 0.1 & $855 \pm 30$ & $7.31 \pm 0.51$ & $1670^{+1813}_{-713}$ & & \\
  & && 1.0  & $85.5 \pm 3.0$ & $0.731 \pm 0.051$ & $303^{+329}_{-129}$ & & 2 \\
 & & & 10.0 & $8.55 \pm 0.30$ & $0.0731 \pm 0.0051$ & $55.1^{+59.8}_{-23.5}$ &  & \\ \hline

 J1901$+$0510  & 429 & 313 & 0.1 & $2145 \pm 35$ & $46.0 \pm 1.5$ & $7560^{+937}_{-780}$ & & \\
 & & & 1.0 & $21.45 \pm 3.5$ & $4.60 \pm 0.15$ & $1370^{+170}_{-145}$ & & 1 \\
 &  & &10.0 & $21.45 \pm 0.35$ & $0.460 \pm 0.015$ & $249^{+31}_{-26}$ & & \\ \hline

  J1907$+$0918   & 358 & 38& 0.1& $1789.5 \pm 0.5$ & $32.023 \pm 0.018$ & $3520^{+3900}_{-2080}$ &  SNR? & \\
& & &1.0 & $178.95 \pm 0.5$ & $3.2023 \pm 0.0018$ & $639^{+711}_{-377}$ & & 5 \\
 & & & 10.0 & $17.895 \pm 0.05$ & $0.32023 \pm 0.00018$ & $116^{+129}_{-68}$ & & \\ \hline

J2007$+$2722  & 127 & 404000 & 0.1 & $635 \pm 2$ & $4.032 \pm 0.025$ & $1140^{+749}_{-446}$ & & \\
  & && 1.0 & $63.5 \pm 0.2$ & $0.4032 \pm 0.0025$ & $207^{+136}_{-81}$ & & 2 \\
 & & & 10.0 & $6.35 \pm 0.02$ & $0.04032 \pm 0.00025$ & $38^{+25}_{-15}$ & & \\ \hline
 \multicolumn{9}{l}{Note: the number in the last unmarked column indicates the number of Figure in which the spectrum, along \phantom{\Large X}}\\
 \multicolumn{9}{l}{ with the fitted model, is plotted. }
\end{tabular}

\label{tabparam}
}
\end{table*}

Following \citet{lewandowski2015a} and \citet{rajwade2016a}, we have to point out that 
the free electrons in the ISM contribute to both the observed dispersion of the pulsar signal as well as to the absorption. In the case of the dispersion effect, the contribution from any region of space is proportional to its electron density $n_e$, but the same region contributes to the optical depth as its $n_e^2$. Using our assumption of a dense uniform region being the source of the absorption, we can estimate its emission measure along the line of sight to be 
$\mathrm{EM}=n_{e}^2 \times s$ , where $s$ is the width of the absorber. On the other, hand this absorber will also contribute to the pulsar's dispersion measure providing $\Delta {\rm{DM}}=n_e \times s$.
 
Our best model fits provide  parameter $B$ which is dependent on the temperature of the electrons and the emission measure. 
Using the above, it can be written as $B = 0.08235 \times T_{\mathrm{e}}^{-1.35}\times n_e^2 \times s$, which gives us the first equation binding these 3 parameters. The dispersion measure provides the second equation, but 
since the DM's dependence on the electron density is not as steep as in the case of thermal absorption, one can not negate the DM contribution from the general ISM (i.e. region outside of the absorber). Obviously, without additional information about the electron density and the size of the absorbing region, we can not reliably estimate the fraction of the total observed DM that comes from the absorber. Following  \citet{rajwade2016a} and \citet{basu2016}, for the purposes of our calculations, we assumed that for the pulsars exhibiting the GPS phenomenon, hence the ones that have a well defined absorber along their line-of-sight, the contribution to the DM from the absorber is equal to half of the observed DM value. This allows us to write the second equation concerning the physical parameters of the absorber: $0.5 \ \mathrm{DM} = n_e\times s$.
 
Since we have only two equations and three free parameters ($s$, $n_e$ and $T_e$), it is impossible to solve for the values of the parameters, without additional information about at least one of them, or without additional assumptions. In the case of pulsar observations the additional information is usually not available, as it is extremely hard to identify the actual absorber, or - in the case of known PWNe - to extract the physical parameters of the region that will contribute to radio wave absorption. Therefore, following \citet{basu2016}, we decided to continue our analysis for three distinct categories of possible absorbers:
\begin{enumerate}
\item dense filaments in an SNR with typical $s = 0.1$ pc,
\item cometary shaped tail in a PWN with $s = 1.0$ pc, and
\item H~II region with $s = 10$ pc.
\end{enumerate}

Table~\ref{tabparam}, in addition to basic and relevant pulsar parameters (i.e. DM, characteristic age and possible associations) shows the results of our calculations for the physical properties of the absorbing region using the three cases described above, i.e. for every pulsar the electron density and temperature was calculated for hypothetical absorbers with physical width of 0.1, 1.0 and 10 parsecs. Based on this information one can attempt to identify the type of absorbing region, by excluding the unphysical or unlikely parameter combinations as explored below.

\subsection{Constrains on possible astrophysical absorbers}

\citet{lewandowski2015a} pointed out that it is extremely difficult to estimate the electron densities and temperatures in the pulsar surroundings, using observational data which are, as a result, extremely rare. In the case of the GPS pulsars identified so far such data are not available. Additionally, the situation is even more futile in the case of pulsar spectral observations, since from such studies one can not ascertain the geometry of the absorber: its actual location and size (i.e. its extent along the line-of-sight) remains unknown. In fact, the location of the absorber is completely irrelevant to the amount of the observed absorption. We can only assume that the absorption happens in the vicinity of the pulsar, since pulsars (especially the young ones), 
are often located in environments with relatively high electron density.

For the above reasons we are, at the moment, unable to predict the actual physical parameters of the absorbing matter that causes the turnover in pulsar spectra, we can, however, put some observational constrains on these absorbers. This allows us, at least to some degree, to distinguish between the different kinds of astrophysical sources of absorption and point out the most likely case, or at the least exclude the non-viable possibilities. 

In Table~\ref{tabparam}, we present the results of our calculations of the physical properties of the absorbers that would cause the observed amount of absorption for each of the GPS pulsars. For each of the sources, we calculated the required electron density and temperature assuming three different sizes of the hypothetical absorber. These represent an absorption in an H~II region (10 pc in size), a shell of a PWN, or a small H~II region (1 pc) and a dense SNR filament (0.1 pc). The corresponding derived values of density and temperature of electrons can be used to estimate the likelihood of a particular type of absorber being the reason of the GPS behavior. For example, in virtually all the cases presented in Table~\ref{tabparam}, the assumption of the H~II absorber yields somewhat realistic values of the electron density, up to few tens of electrons per cubic centimeter, however the corresponding electron temperatures are extremely low (below 100~K) which is unphysical for this type of ionized region (typical electron temperatures in H~II regions are of the order of a few thousand kelvins). Therefore we can confidently exclude them as possible absorbers.

In general, we believe that our calculations clearly show that extended H~II regions are the most unlikely causes for the appearance of the GPS spectra. The high electron temperature in such regions causes the ISM to be more transparent. To provide enough absorption in such a case the electron density would have to be much higher. While the high value of density on its own is not an issue, since  some of the H~II regions exhibit densities up to $10^5$ particles per cm$^3$, one has to realize that the same region would also provide the dispersion of the pulsar signal. The value of the observed DM is easy to calculate: a 10 pc region with an electron density of 1000 particles per cm$^3$ would provide 10000~pc~cm$^{-3}$ of dispersion, making the pulsar very unlikely, or even impossible, to be discovered in a regular pulsar survey (even provided that the survey would be conducted at high frequency, where  the thermal absorption would not play a significant role).

As for the other types of absorbers, the interpretation is usually not as straightforward, since these kinds of pulsar environments can exhibit a range of physical parameters. As  \citet{lewandowski2015a} pointed out, the dense filaments in a supernova remnant can be dense enough (up to a few thousand particles per cubic centimeter) that, even when considering their high temperature (up to a few thousand kelvins), they still provide enough absorption to cause spectral turnover at gigahertz frequencies, while being only a fraction of a parsec in size. Looking at the data in Table~\ref{tabparam} a number of pulsars seems to fit the criteria: for example PSR~B1054$-$62 ($n_e = 1601$ cm$^{-3}$, $T_e=5080 K$), or PSR~B1822$-$14 ($n_e = 1785$ cm$^{-3}$, $T_e=10100 K$), and several others. Obviously, not all of the GPS pulsars listed in the table that would fit the SNR filament scenario were actually observed within an SNR, however this can be caused by the fact that  the distant remnants are especially difficult to detect. Additionally, one has to remember that the absorber does not have to be physically or evolutionary connected to the remnant - just like in the case of PSR~B1800$-$21, studied by \citet{basu2016}, an object whose line-of-sight apparently crosses an unrelated W30 remnant nebula.

Distinguishing between absorption in a dense filament and absorption in a shell (or rather ``tail'') of a pulsar wind nebula may be more difficult.  \citet{lewandowski2015a} showed that the measurements of the temperatures and densities in cometary-shaped PWNe are very sparse, at least when it comes to these parts of the nebulae that could possibly provide free-free absorption in the radio regime. They also show that if one considers densities of the order of a few hundred particles per cm$^3$, then for roughly 1pc size of the absorber in the PWN tail, one would need temperatures as low as few hundred kelvins, to provide significant absorption at gigahertz frequencies. Based on the data we show in Table~\ref{tabparam}, several pulsars would qualify in this category, like for example 
PSR~B1641$-$45 ($n_e=239$~cm$^{-3}$, $T_e =983$~K). However, for these sources the SNR filament scenario also provides reasonable parameters, making it extremely difficult to decide which of these two possibilities we are dealing with for a particular pulsar. To solve this one would need some additional observations, like the detection of a PWN, or measurements of the associated/coincident SNR, however in the case of the confirmed GPS pulsars such observations are not available, and even in the cases where we have a confirmation of a pulsar wind nebula or supernova remnant - the data are not detailed enough to provide information that would allow us to estimate the parameters relevant to the thermal absorption of radio waves. Finally, since the effect of absorption in our model does not depend on the actual location of the absorber, we expect that in some cases we may see the pulsars through absorbers that are not in their immediate vicinity, which makes the interpretation even harder; the possibility of multiple absorbers 
 (with different physical parameters) cannot also be excluded.

\subsection{Finding GPS pulsars in future surveys}
\label{surveys}

Based on the values of peak frequency we obtained (see $\nu_p$ in Table~\ref{fitparams}) we created a histogram of the peak frequency distribution which is shown in Figure~\ref{fig8}. To create the histogram we included all the pulsars and radio-magnetars presented in this work, as well as the data from the binary pulsar B1259$-$63 \citep[][we used the lowest peak frequency found there, 1.6~GHz]{kijak2011a}, the Galactic center radio-magnetar J1745$-$2900 \citep[][the peak frequency slightly larger than 2~GHz]{pennucci2015}  and PSR~B1800$-$21 \citep{basu2016}.
While the statistics is still small, looking at the histogram one can note a difference between the regular GPS pulsars and the peak frequencies in the spectra of radio magnetars. This may be at least partially explained by the differences in the methods of discovery. The radio-magnetars are always found by targeted searches following an X-ray outburst of the source, and the X-ray data often provides the rotational period, which significantly narrows the parameter-space that needs to be searched to find  the radio counterpart. 

On the other hand finding a GPS pulsar with a peak frequency higher than, say, 1.5 GHz will be difficult. An optimal frequency range for a GPS pulsar search is definitely much higher than its peak frequency. For example, if $\nu_{0.9}$  is the frequency at which one would see 90\% of the pulsar's flux, then using simple algebra one can show that $\nu_{0.9}/\nu_p \approx \sqrt{4.5 \times \alpha}$ (where $\alpha$ is the intrinsic pulsar spectral index). For a typical pulsar ($\alpha \approx 2$) it means that only in the frequency range above 3 times the peak frequency the observer receives almost all of the pulsar intrinsic flux, making this range optimal for detection. This would explain the lack of regular pulsars with  $\nu_p$ above 2~GHz, since the search for such sources would be most effective at frequencies higher than 6~GHz. 
There were only a few limited attempts to search for pulsars in this range, like for example the Parkes Methanol Multibeam Survey \citep{bates2011}. One of the main reasons why such searches are not attempted more often is that blind surveys at such high frequencies are extremely time consuming. This is due to the decreased telescope beam size, which in turn decreases the Galaxy volume that may be searched in a given time. Additionally, GPS pulsars with such a high absorption would also most likely exhibit high values of DM, and that would happen regardless of their distance, since the absorption and its associated high DM contribution is most probably local to the pulsar. A possibility of a large DM value may significantly increase the parameter space that has to be searched, however, since the dispersion smearing is inversely proportional to the observing frequency squared this may not be a big issue for such high frequency searches.

\citet{rajwade2016b} recently published a detailed study of past and proposed high frequency pulsar surveys, calculating the expected discovery probabilities while taking into account the effects of thermal free-free absorption and  scattering. They show that the optimal frequency range for finding pulsars in the central regions of the Milky Way would be 9 to 13~GHz, and they claim that the main limiting factor in this case will be rather the interstellar scattering, not absorption. However, in their calculation of thermal absorption they used an emission measure ($EM$) of $5\times 10^5$~pc~cm$^{-6}$ for the Galactic Center pulsars. As we show in Table~\ref{tabparam} in some scenarios the GPS spectrum may be explained by regions with $EM$ exceeding $10^6$~pc~cm$^{-6}$ (see the radio-magnetars in the  0.1 pc absorber case), which means that for very dense environments the thermal absorption may play much stronger role in the probability of detection, than \citet{rajwade2016b} used for the proposed Galactic Center surveys.

For the above reasons we believe, that there is still a high chance that we can discover GPS pulsars in some of the less studied PWNe or SNRs, by the means of targeted searches at frequencies close to and above 10~GHz. Additional targets for such searches may be the unresolved X-ray sources that show characteristics of a PWN (for example with regard to their spectra). Also, there is a chance that a number of GPS pulsars still remains undetected in the larger/more dense/cooler H~II regions, and possibly in the center of the Milky Way. Blind surveys performed at higher frequencies may also yield occasional discoveries, however they will be much more time consuming.

\subsection{GPS radio magnetars}

We also included in our studies two radio magnetars, J1550$-$5418 and J1622$-$4950 that were previously reported by \citet{kijak2013} to exhibit GPS type spectra, with the apparent peaks at very high frequencies. Another case was studied by  \citet{lewandowski2015a}, namely the Sgr~A* radio magnetar, located very close to the central black hole of our Galaxy, and this object was also exhibiting GPS characteristics, which evolved with time - since, according to the model shown by the authors, the amount of absorption was decreasing with time.

One has to note a significant difference between our results and the study of the  Sgr~A* magnetar: the spectra that were analysed for this object were obtained for individual epochs, while the spectra we show here for  J1550$-$5418 and J1622$-$4950 were obtained from all available radio flux density measurements, regardless of the date of observations. Therefore a certain degree of caution is required when interpreting the results. If the radio spectra of the magnetars indeed change over time, the spectra we used for our models do not correspond to any given moment in time and their shape may be heavily affected by the actual evolution. For that reason, the physical parameters inferred from the models should be treated only as rough/average  estimates.

\citet{kijak2013} noted that the peak frequencies for the two radio magnetars are much higher than the values obtained for the remaining GPS pulsars, suggesting that the magnetars have to be in even more extreme environments compared to the other GPS pulsars.  This would be explained by the fact, that these magnetars are extremely young objects (see Table~\ref{tabparam}, the ages of 1410 and 4030 years), which would indicate that one can expect much more extreme conditions in their surroundings, especially when it comes to the electron density. The density influence will be somewhat offset by higher temperatures of these surroundings, however since the dependence on density (through the emission measure) is much stronger, that would explain higher amounts of absorption and higher peak frequencies. We also have to note that
the peak frequencies obtained from our thermal absorption fits are significantly lower than the ones reported before, i.e. 3.27 and 3.7~GHz instead of 5.0 and 8.3~GHz (see also Fig.~\ref{fig8}). This discrepancy comes from the fact that the previous paper used the purely morphological model of parabolic spectral shape proposed by \citet{kuzmin2000}. The application of the real physical model suggests that the radio magnetars have relatively flat spectra: the intrinsic spectral indexes we obtained from our modeling are the lowest in the sample ($-0.46$ and $-0.54$) which combined with the thermal absorption profile causes only very small changes in the observed flux density over a relatively wide range 
 of frequencies. The use of the actual model moves the peak frequency from the middle of that ``almost flat'' spectral region towards  lower frequencies, yielding the values around 3~GHz.  \citet{lewandowski2015a} in their study of the Sgr~A* radio magnetar obtained a similar value (which can be inferred from their spectra plots),  about 3.5~GHz at 40 days since the magnetar outburst. Also \citet{pennucci2015}  observed the spectrum of that magnetar, and their plots suggest a peak frequency of about 2 to 2.5~GHz, and their observations were made approximately 100 days after the outburst. All of these studies clearly indicate that if the thermal absorption is the cause of the spectral turnovers in radio magnetars, then indeed the parameters of the absorbers must be more extreme than in the case of regular GPS pulsars.

In Section~\ref{surveys} we discussed some basic aspects of GPS pulsar searches. Based on that we believe, that the main reason for the fact that we know of magnetars with such high peak frequencies, but we did not discover pulsars exhibiting such high amount of absorption is purely due to the discovery bias: the magnetars are discovered in the radio regime only after their initial X-ray outburst discovery (which often provides the rotational parameters as well), while for regular pulsars we do not have that option and we are forced to perform a full search, which (as we discussed above) can succeed only if performed at sufficiently high observing frequency. It is possible, that there are pulsars located in similarly extreme environments that simply were not discovered yet, because no one searched for them at observing frequencies around 10~GHz and above. 

Moreover, there is a possibility that the reason that some of the magnetars do not exhibit radio emission at all may be due to thermal absorption that is even more extreme than in the cases of  J1550$-$5418 and J1622$-$4950. However, given the nearly flat intrinsic radio-magnetar spectra, these objects would be easier to find at frequencies around and above 10~GHz, and we know that such attempts were made, at least in some selected cases. Hence we believe, that the free-free absorption as the explanation for radio-quiet magnetars is not very likely.

\section{Summary}
We used the interferometric imaging technique to estimate the low radio
frequency flux in fifteen pulsars. The high sensitivity of the measurements 
allowed us to construct the spectral shape of these pulsars and in the process 
we identified five new pulsars with gigahertz-peaked spectra (GPS). 
Additionally, our measurements resulted in tighter constrains on the low 
frequency spectra of four pulsars and confirmed their GPS characteristics. 
Summarizing, the GPS pulsar population now consists of seventeen sources.

The GPS phenomenon in pulsars is most likely a result of thermal absorption of 
the pulsar flux in specialized environments along the line of sight. A detailed
study was conducted where the low frequency turnover in the spectral shape was
successfully modeled using thermal absorption, using 3-parameter fit procedure similar to the one
employed by \citet{basu2016}. We presented the results of our modeling for fourteen GPS pulsars
(see Figs. \ref{newGPS}, \ref{fig2GPS}, \ref{fig5}, \ref{fig6}). 
The remaining three GPS neutron stars:   PSR~B1800$-$21,
PSR~B1259$-$63~in the binary system with Be star LS 2883 and SGR~J1745$-$2900 have been modeled in separate works (see section 4.4). A detailed study of PSR~J1740+1000, a source which was claimed to show GPS characteristics by \citet{dembska2014}, is in progress. 
The spectral shape arising due to the thermal absorption is not sufficient to fully
characterize the physical properties like temperature, electron density and 
size of the absorbing region. However, using basic physical arguments, we have 
explored the nature of the absorber and ruled out some of the potential candidates.

We also discussed the strategies for finding GPS pulsars in the future search surveys. The optimal frequency range  is usually a few times larger than the peak frequency, meaning that for a normal GPS pulsar (with a peak frequency close to 1~GHz) the optimal frequency would be greater than 3~GHz, while for the sources with much stronger absorption, similar to radio-magnetars ($\nu_p$ close to 3~GHz), the optimal frequency would be much higher, possibly close to 10~GHz. Therefore we believe that a targeted search surveys of known PWNe, H~II regions and supernova remnants should be the way to go. However, we have to also  point out that  there is  a possibility, that a lot of GPS sources still hides in the known pulsar population. Such sources were not identified yet simply because for the majority of pulsars the shapes of their spectra remain unknown, especially in the low frequency range (below 1~GHz) where thermal free-free absorption would manifest itself.

\section*{Acknowledgments}
\acknowledgments{
We thank the staff of the GMRT who have made these observations possible. 
The GMRT is run by the National Centre for Radio Astrophysics of the Tata 
Institute of Fundamental Research. This research was partially supported by the
grant  DEC-2013/09/B/ST9/02177 of the Polish National Science Centre. }


\begin{thebibliography}{99}
%\bibitem[\protect\citeauthoryear{Aharonian et al.}{2005}]{ah1}Aharonian, F. et al. 2005, Science, 307, 1938
%\bibitem[\protect\citeauthoryear{Aharonian et al.}{2006}]{ah2} Aharonian, F. et al. 2006, Ap. J. 636, 777


\bibitem[\protect\citeauthoryear{Allen et al.}{2013}]{allen2013} Allen, B., Knispel, B., Cordes, J.~M., et al. 2013, ApJ, 773, 91 (A13)

\bibitem[\protect\citeauthoryear{Anderson et al.}{2012}]{anderson2012} Anderson, G. E., Gaensler, B. M., Slane, P. O., et al. 2012, ApJ, 751, 53


\bibitem[\protect\citeauthoryear{Bates et al.}{2011}]{bates2011} Bates, S. D., Johnston, S., Lorimer, D. R., et al. 2011, MNRAS, 411, 1575 (B11)

%\bibitem[\protect\citeauthoryear{Baars et al.}{1977}]{baars} Baars, J. W. M., Genzel, R., Pauliny-Toth, I. I. K., Witzel, A. 1977, A\&A, 61, 99
%\bibitem[\protect\citeauthoryear{Basu et al.}{2011}]{basu1}Basu R., Athreya R., Mitra D. 2011, ApJ, 728, 157
%\bibitem[\protect\citeauthoryear{Basu et al.}{2012}]{basu2}Basu R., Mitra D., Athreya R. 2012, ApJ, 758, 91

\bibitem[\protect\citeauthoryear{Basu et al.}{2016}]{basu2016}Basu, R., Ro\.zko, K., Lewandowski, W., Kijak, J., \& Dembska, M. 2016, MNRAS, 458, 2509

\bibitem[\protect\citeauthoryear{Bates et al.}{2013}]{bates2013}Bates S. D., Lorimer D. R., \& Verbiest J. P. W. 2013, MNRAS, 431, 1352

%\bibitem[\protect\citeauthoryear{Bhat et al.}{2004}]{bhat2004}Bhat N. D. R., Cordes J.M., Camillo F. et al. 2004, ApJ, 605, 759

\bibitem[\protect\citeauthoryear{Camilo et al.}{2007}]{camilo2007} Camilo, F., Ransom, S. M., Halpern, J. P., \& Reynolds, J. 2007, ApJL, 666, L93

\bibitem[\protect\citeauthoryear{Camilo et al.}{2008}]{camilo2008}Camilo, F., Reynolds, J., Johnston, S., Halpern, J. P., \& Ransom, S.M. 2008, ApJ, 679, 681

\bibitem[\protect\citeauthoryear{Champion et al.}{2005}]{champion2005} Champion, D. J., Lorimer, D. R., McLaughlin, M. A., et al. 2005,  MNRAS, 363, 929 (C05)

\bibitem[\protect\citeauthoryear{Costa et al.}{1991}]{costa1991} Costa, M. E.,McCulloch, P.M.,  \& Hamilton, P. A. 1991, MNRAS, 252, 13 (C91)


\bibitem[\protect\citeauthoryear{Dembska et al.}{2014}]{dembska2014} Dembska M., Kijak J.,  Lewandowski W., Jessner A., Bhattacharyya B., \& Gupta Y. 2014, MNRAS, 445, 3105 (D14)
%\emph{accepted for publications in MNRAS}

\bibitem[\protect\citeauthoryear{Dembska et al.}{2015a}]{dembska2015a} Dembska M., Basu R., Kijak J., \& Lewandowski W. 2015a, MNRAS, 449, 1869 (D15)

\bibitem[\protect\citeauthoryear{Dembska et al.}{2015b}]{dembska2015b} Dembska M., Kijak J., Koralewska O., Lewandowski W., Melikidze G., \& Ro\.zko K. 2015b, Ap\&SS, 359, 31 

\bibitem[\protect\citeauthoryear{Hobbs et al.}{2004}]{hobbs2004} Hobbs, G., Faulkner, A., Stairs, I. H., et al. 2004, MNRAS, 352, 1439 (H04)

\bibitem[\protect\citeauthoryear{Johnston et al.}{2006}]{johnston2006} Johnston, Simon., Karastergiou, Aris., \& Willett, Kyle. 2006, MNRAS, 369, 1916 (J06)

%\bibitem[\protect\citeauthoryear{Finley \& Oegelman}{1994}]{fo}Finley J.~P., Oegelman, H., ApJ, 434, L25
\bibitem[\protect\citeauthoryear{Geyer \& Karastergiou}{2016}]{geyer2016} Geyer, M., Karastergiou, A., 2016, MNRAS, 462, 2587
%\bibitem[\protect\citeauthoryear{Higashi et al.}{2008}]{hig}Higashi, Y. et al. 2008, ApJ 683, 957
%\bibitem[\protect\citeauthoryear{Kassim \& Weiler}{1990a}]{kw1}Kassim, N. E., Weiler, K. W. 1990a, Nature, 343, 146
%\bibitem[\protect\citeauthoryear{Kassim \& Weiler}{1990b}]{kw2}Kassim, N. E., Weiler, K. W. 1990b, ApJ, 360, 184

\bibitem[\protect\citeauthoryear{Keith et al.}{2011}]{keith2011} Keith, M. J., Johnston, S., Levin, L.,  \& Bailes, M. 2011, MNRAS, 416, 346 (KJL11)

\bibitem[\protect\citeauthoryear{Kijak et al.}{1998}]{kijak1998} Kijak, J., Kramer, M., Wielebinski, R.,  \& Jessner, A. 1998, A\&AS, 127, 153 (K98)

\bibitem[\protect\citeauthoryear{Kijak et al.}{2004}]{kijak2004} Kijak, J., \& Maron, O. 2004, PASPS, 218, 339 

\bibitem[\protect\citeauthoryear{Kijak et al.}{2007}]{kijak2007} Kijak J., Gupta Y., \& Krzeszowski K. 2007, A\&A, 462, 699 (K07)

\bibitem[\protect\citeauthoryear{Kijak et al.}{2011a}]{kijak2011a} Kijak, J., Dembska, M., Lewandowski, W., Melikidze, G., \& Sendyk, M. 2011a, MNRAS, 418, L114

\bibitem[\protect\citeauthoryear{Kijak et al.}{2011b}]{kijak2011b} Kijak, J., Lewandowski, W., Maron, O., Gupta, Y.,  \& Jessner, A. 2011b, A\&A, 531, A16 (K11)

\bibitem[\protect\citeauthoryear{Kijak et al.}{2013}]{kijak2013}Kijak, J., Tarczewski, L., Lewandowski, W., \& Melikidze, G. 2013, ApJ, 772, 29

%\bibitem[\protect\citeauthoryear{Kouwenhoven}{2000}]{kouwenhoven}Kouwenhoven M. L. A. 2000, A\&AS, 145, 243

\bibitem[\protect\citeauthoryear{Kuzmin \& Losovsky}{2000}]{kuzmin2000} Kuzmin, A.D., \& Losovsky, B.Ya. 2000, Astronomy Letters, 26, 500

\bibitem[\protect\citeauthoryear{Krishnakumar~et~al.}{2015}]{krishna15}
Krishnakumar, M.A., Mitra, D., Naidu, A., Joshi, B.C. \& Manoharan P.K. 2015, ApJ, 804, id:23

\bibitem[\protect\citeauthoryear{Levin et al.}{2010}]{levin2010} Levin, L., Bailes, M., Bates, S., et al. 2010, ApJL, 721, L33

\bibitem[\protect\citeauthoryear{Levin et al.}{2012}]{levin2012} Levin, L., Bailes, M., Bates, S. D., et al. 2012, MNRAS, 422, 2489

\bibitem[\protect\citeauthoryear{Lewandowski et al.}{2013}]{lewandowski2013} Lewandowski, W., Dembska, M., Kijak, J., \& Kowali\'nska, M. 2013, MNRAS, 434, 69

\bibitem[\protect\citeauthoryear{Lewandowski et al.}{2015a}]{lewandowski2015a}  Lewandowski, W., Ro\.zko, K.,  Kijak, J., \& Melikidze, G.I. 2015a, ApJ, 808, id:18

\bibitem[\protect\citeauthoryear{Lewandowski, Kowali\'nska \& Kijak}{2015b}]{lewandowski2015b}  Lewandowski, W., Kowalinska, M., \& Kijak, J. 2015b, MNRAS, 449, 1570

\bibitem[\protect\citeauthoryear{Lewandowski et al.}{2015c}]{lewandowski2015c}  Lewandowski, W., Ro\.zko, K., Kijak, J., Bhattacharyya, B., \& Roy, J. 2015c, MNRAS, 454, 2517


\bibitem[\protect\citeauthoryear{Lorimer et al.}{1995}]{lorimer1995} Lorimer D. R., Yates J. A., Lyne A. G., \& Gould D. M. 1995, MNRAS, 273, 411 (L95)

%\bibitem[\protect\citeauthoryear{Lorimer et al.}{1995}]{lorimer95} Lorimer, D. R., Yates, J. A., Lyne, A. G.,  Gould, D. M., 1995, MNRAS, 273, 411

\bibitem[\protect\citeauthoryear{Lorimer et al.}{2000}]{lorimer2000} Lorimer, Duncan R., \& Xilouris, Kiriaki M. 2000, ApJ, 545, 385L (L00)

\bibitem[\protect\citeauthoryear{Manchester et al.}{1978}]{manchester1978} Manchester, R. N., Lyne, A. G., Taylor, J. H., Durdin, J. M., Large, M. I., \& Little, A. G. 1978, MNRAS, 185, 09 (ML78)

\bibitem[\protect\citeauthoryear{Manchester et al.}{2001}]{manchester2001} Manchester, R. N., Lyne, A. G., Camilo, F., et al. 2001,  MNRAS, 328, 17 (ML01)


\bibitem[\protect\citeauthoryear{Manchester et al.}{2005}]{manchester2005} Manchester, R. N., Hobbs, G. B., Teoh, A.,  \& Hobbs, M. 2005, AJ, 129, 1993

\bibitem[\protect\citeauthoryear{Maron et al.}{2000}]{maron2000} Maron O., Kijak J., Kramer M., \& Wielebinski R. 2000, A\&AS,147, 195 (M00)

\bibitem[\protect\citeauthoryear{McLaughlin et al.}{2002}]{mcl2002} 	McLaughlin, M. A., Arzoumanian, Z.,  Cordes, J. M., Backer, D. C., Lommen, A. N., Lorimer, D. R., \& Zepka, A. F. 2002, ApJ, 564, 333

\bibitem[\protect\citeauthoryear{Morris et al.}{2002}]{morris2002} Morris, D. J., Hobbs, G., Lyne, A. G., et al. 2002,  MNRAS, 335, 275 (MH02)

\bibitem[\protect\citeauthoryear{Pennucci et al.}{2015}]{pennucci2015} Pennucci, T.T., Possenti, A., Esposito, P., Rea, N., Haggard, D., Baganoff, F.K., Burgay, M., Coti Zelati, F., Israel, G.L., \& Minter, A. 2015, ApJ, 808, 81

\bibitem[\protect\citeauthoryear{Perley \& Butler}{2013}]{perley2013}Perley, R. A., \& Butler, B. J. 2013, ApJS, 204, 19

\bibitem[\protect\citeauthoryear{Rajwade, Lorimer \& Anderson}{2016a}]{rajwade2016a} Rajwade, K., Lorimer, D.R., \& Anderson, L.D. 2016a, MNRAS, 455, 493

\bibitem[\protect\citeauthoryear{Rajwade, Lorimer \& Anderson}{2016b}]{rajwade2016b} 
Rajwade, Kaustubh, Lorimer, Duncan, Anderson, Loren 2016b,  arXiv:1611.06977


\bibitem[\protect\citeauthoryear{Rybicki \& Lightman}{1979}]{RyLa79} Rybicki, G.B. \& Lightman, A.P. 1979, \textit{Radiative Processes in astrophysics}, New York, Chichester, Brisbane, Toronto, Singapore: John Wiley \& Sons, Inc.

\bibitem[\protect\citeauthoryear{Sieber}{1973}]{sie73}Sieber, W. 1973, A\&A, 28, 237

%\bibitem[\protect\citeauthoryear{Swarup et al.}{1991}]{swarup1991} Swarup, G., Ananthakrishnan, S., Kapahi, V. K., Rao, A. P., Subrahmanya, C. R., Kulkarni, V. K.  1991, CuSc, 60, 95 

\bibitem[\protect\citeauthoryear{Taylor et al.}{1993}]{taylor1993} Taylor, J. H., Manchester, R. N., \& Lyne, A. G. 1993, ApJS, 88, 529 (T93)

\bibitem[\protect\citeauthoryear{van Ommen et al.}{1997}]{ommen1997} van Ommen, T. D., D’Alessandro, F., Hamilton, P. A., \& McCulloch, P.M. 1997, MNRAS, 287, 307 (vO97)

\bibitem[\protect\citeauthoryear{Wilson et al.}{2009}]{Wil2009} Wilson, T.~L., \& Rohlfs, K., H\"{u}ttemeister S. 2009, \textit{Tools of Radio Astronomy}, Berlin: Springer 

\bibitem[\protect\citeauthoryear{Wu et al.}{1993}]{wu1993} Wu X., Manchester, R. N., Lyne, A. G., \&  Qiao, G. 1993, MNRAS, 261, 630 (W93)

\bibitem[\protect\citeauthoryear{Young et al.}{2015}]{young2015} Young, N.J., Weltevrede, P., Stappers, B.W., Lyne, A.G., \& Kramer, M. 2015, MNRAS, 449, 1495

\end{thebibliography}
\end{document}